\documentclass{birkjour}

\usepackage{cite,enumitem}
\usepackage{graphicx}

\theoremstyle{plain}
\newtheorem{Theorem}{Theorem}
\newtheorem{Lemma}{Lemma}
\newtheorem{Proposition}{Proposition}
\newtheorem{Corollary}{Corollary}
\theoremstyle{remark}
\newtheorem*{Remark}{Remark}
\newtheorem{examp}{Example}
\newtheorem{exampB}{Example}

\newcommand{\bpow}{\mathbf{p}}
\newcommand{\tr}{\operatorname{tr}}

\newcommand{\di}{\mathrm{dim}}

\def\be{\begin{equation}}
\def\ee{\end{equation}}
\def\bl{\begin{Lemma}}
\def\el{\end{Lemma}}

\begin{document}
	
	\author{Aleksander Yu. Orlov}
	\address{Institute of Oceanology\\
		Nahimovskii Prospekt 36\\
		Moscow 117997, Russia\\
		and\\
		National Research University Higher School of Economics\\
		International Laboratory of Representation\\
		Theory and Mathematical Physics\\
		20 Myasnitskaya Ulitsa\\
		Moscow 101000, Russia}
	\email{orlovs@ocean.ru}
	\title[Hurwitz numbers and matrix integrals labeled with chord diagrams]
	{Hurwitz numbers and matrix integrals
	\\  $\qquad$ labeled with chord diagrams}
	
	\keywords{Hurwitz number, Schur functions, Klein surface, independent complex Ginibre ensembles,
		products of complex random matrices, Euler characteristic of network chord diagrams, 
		ribbon graph, gluing of polygons, discrete beta-ensembles}
	
	\subjclass{05A15, 14N10, 17B80, 35Q51, 35Q53, 35Q55, 37K20, 37K30}

	\begin{abstract}
	
We consider products of complex random matrices from independent complex Ginibre ensembles.
The products include complex random matrices $ Z_i, Z_i^\dag, \, i = 1, \dots, n $, and $ 2n $ sources
(these are the complex matrices $ C_i, C_i^*, \, i = 1, \dots, n $, which play the role of parameters).
We consider collections of products $ X_1, \dots, X_F $, constainted by the property,
that each of the matrices of the set $\{ Z_iC_i, Z_i^\dag C_i^*,\,i=1,\dots, n \}$ is included 
only once on the product $ X = X_1 \cdots X_F $.
It can be represented graphically as a collection of $ F $ polygons with a total number of edges $ 2n $,
and the polygon with number $ a $ encodes the order of the matrices in $ X_a $.
The matrices $ Z_i $ and $ Z_i^\dag $ are distributed along the edges of this collection of polygons, and
the sources are distributed at their vertices. The calculation of the expected values involves pairing the 
matrices $ Z_i $ and $ Z_i^\dag $.
There is a standard procedure for constructing a $ 2D $ surface by paiwise gluing edges of polygons,
this procedure results to a {\it ribbon graph} embedded in the surface $\Sigma_{\textsc{e}^*}$ of some Euler characteristic 
$\textsc{e}^*$
(this graph also known as {\it embedded graph} or {\it fatgraph}). We propose a matrix model that generates spectral 
correlation functions for matrices $ X_a, \, a = 1, \dots, F $ in the Ginibre ensembles, which we call the 
matrix integral, labeled network chord diagram. We show that the spectral correlation functions generate
Hurwitz numbers $H_{\textsc{e}^*}$ that enumerate nonequivalent branched coverings of  $ \Sigma_{\textsc{e}^*} $.
The role of sources is the generation of ramification profiles in branch points which are assigned to the 
vertices of the ribbon graph drawn on the base surface $\Sigma_{\textsc{e}^*}$. The role of coupling constants 
of our model is to generate ramification profiles in $F$ additional branch points assigned to the faces of 
the ribbon graph (the faces of the 'triangulated' $\Sigma_{\textsc{e}^*}$).
 The Hurwitz numbers for Klein surfaces can also be obtained by a small modification of the model. 
 To do this, we pair any of the source matrices (in that case presenting a hole on $\Sigma_{\textsc{e}^*}$) 
 with the tau function, which we call Mobius one.
The presented matrix models generate Hurwitz numbers for any given Euler
characteristic of the base surface $ \textsc{e}^* $ and for any given set of ramification profiles.

\end{abstract}

\maketitle

\section{Introduction \label{Introduction}}

Hurwitz numbers count $d$-sheeted branched covers of Riemann surfaces of a given genus (we will denote it
$g^*$), see for instance \cite{ZL} for a review. The direct analogue exits also for the case of Klein 
surfaces, see \cite{AN},\cite{AN2008}.
A number of facts is known for the topic of Hurwitz numbers and matrix integrals, see 
\cite{MelloKochRamgoolam},\cite{AMMN-2011},\cite{Alexandrov},\cite{KZ},\cite{AMMN-2014},
\cite{AlexandrovZabrodin-Okounkov},\cite{GGPN},
\cite{Harnad-2014},\cite{NO-2014},
\cite{Chekhov-2014},\cite{Chekhov-2016},\cite{NO-LMP}, \cite{Harnad-overview-2015},
\cite{ChekhovAmbjorn}. On the other hand, last few years products of random matrices were in the focus
of studies for applications in quantum chaos and in information theory \cite{Ak1},\cite{Ak2},
\cite{AkStrahov},\cite{S1},\cite{S2}. The relation of these two topics was considered  in \cite{OrlovStrahov},
\cite{O-TMP-2017}. In \cite{OrlovStrahov} it was shown that the partition function of the matrix model
generating spectral correlators for the product of complex
matrices is the Toda lattice \cite{UT} (see also \cite{Takasaki2018} for the overview) tau function of the type
introduced in \cite{OS-2000} (earlier appeared in different form in \cite{KMMM}). 
Let us note that the relation of a number of matrix models with Hurwitz numbers follows directly from
comparing results of \cite{Goulden-Jackson-2008} and of \cite{O-2004-New} (see also 
\cite{HO-2006}).
In \cite{O-TMP-2017} special products of complex matrices from independent Ginibre ensembles 
were considered to generate
Hurwitz numbers in case the Euler characteristic of the base surface is less than 2.
Here, we develope \cite{O-TMP-2017} for the case of any given product of complex matrices
which is suitable to encode by chord diagrams. 
We show that the related matrix integral generates a discrete $\beta$-ensemble (where
the Euler characteristic of the base surface plays the role of $\beta$), thus, instead
of integration over $nN^2$ complex variables we get summation over $N$ variables (this may
be compared to \cite{OShiota-2004} and \cite{Alexandrov}), see formulae
(\ref{Th-1-in-Schur-series}),(\ref{Theor-1-Klein-Schur}),(\ref{Th-1-Schur-U}). 

The present work does not deal with the study of Hurwitz numbers in the framework of integrable systems
which was started in the pioneer works of Okounkov 
\cite{Okounkov-2000,Okounkov-Pand-2006} and later in the article by Goulden and Jackson
\cite{Goulden-Jackson-2008} which was further developed in many papers\footnote{ see
\cite{MM1},\cite{MM2},\cite{AMMN-2011},\cite{AMMN-2014},\cite{Harnad-2014},\cite{NO-2014}
,\cite{HO-2014},\cite{Zog},\cite{Carrell},\cite{DuninKazarian---2013},\cite{NZ},\cite{NO-LMP} 
(see also reviews \cite{Harnad-overview-2015} and \cite{Uspehi-KazarianLando}). 
where the hypergeometric Toda lattice tau functions were used to enumerate covers
of the Riemann sphere, and also \cite{NO-2014},
\cite{NO-LMP} where hypergeometric BKP tau functions were used to enumarate covers of the real 
projective plane $\mathbb{RP}^2$.}

A brief summary of the present work is presented in the Abstract.

\section{Preliminaries}

\subsection{Hurwitz numbers \label{Hurwitz numbers}}

The geometric definition of Hurwitz numbers can be found in the Appendix \ref{Counting-of-branched-covers}. 
Here we give combinatorial definition.

\paragraph{Orientable case.}
Consider symmetric group $S_d$ and the equation
\be\label{Hom-pi-S_d-Riemann}
A_1\cdots A_k
\prod_{j=1}^{g^*} a_jb_ja_j^{-1}b_j^{-1} =1
\ee
where $a_j,b_j,A_i \in S_d,\,j=1,\dots,g^*,\,i=1,\dots,k$ and where each $A_i$ belongs to a given
cycle class $\textsc{C}_{\Delta^i}$. Then the number of the solutions of this equation over the order of the 
symmetric group 
\be
H_{2-2g^*}(\Delta^1,\dots,\Delta^k)=\# \frac{1}{d!}
\{a_1,\dots,a_{g^*},b_1,\dots,b_{g^*},A_1,\dots,A_k \in S_d| A_i\in \textsc{C}_{\Delta^i}  \}
\ee
is called Hurwitz number. These numbers admit geometrical intepretation. In short the Hurwitz
number enumerate branched $d$-sheeted covers of a Riemann surface $\Sigma_{g^*}$ of genus $g^*$
by (not necceraly connected) Riemann surfaces
with given ramification profiles $\Delta^i,\,i=1,\dots,k$
at each of $k$ critical points,  
details may be found in the Appendix. The genus $g$ of a cover $\Sigma_g$ is defined with the help
of the Riemann-Hurwitz relation
\be\label{RHur}
\textsc{e}^*=d\textsc{e}-\sum_{i=1}^k(d-\ell(\Delta^i))
\ee 
where $\ell(\Delta^i)$ is the length of the partition $\Delta^i$ and
where $\textsc{e}^*$ and $\textsc{e}$ are Euler characteristics respectively of the base and of the 
cover (respectively
equal to $2-2g^*$ and to $2-2g$). 
In the geometric interpretation equation (\ref{Hom-pi-S_d-Riemann}) results from the 
 homomorphism of the fundamental group of the (base) Riemann surface $\Sigma_{g^*}$ to the symmetric
 group which acts on the numbered $d$ shieves of the cover. The path around a critical points, say, 
 $z_i\in\Sigma_{g^*}$
 maps to the product of the cyclic permutations related to the ramification profile $\Delta^i$.

$$
\,
$$

\paragraph{Non-orientable case.} 
The enumeration problem of counting of branched $d$-sheeted coverings of 
 Klein surfaces of the Euler characteristic $\textsc{e}^*=2-\textsl{g}^*$ by other Klein (or Riemann)
surfaces\footnote{It is important
that in this consideration only isolated critical points are admissible.} may be reduced to the counting 
of the number of the solutions of the equation
\be\label{Hom-pi-S_d-Klein}
A_1\cdots A_{k}
\prod_{j=1}^{\textsl{g}^*} R_j^2  =1
\ee
where $R_j,A_i\in S_d$ and where each $A_i \in C_{\Delta^i}$, where $\Delta^i,\,i=1,\dots,k$
are the set of given ramification profiles in the critical points.  Similarly to the previous case,
Hurwitz number may be defined as the number of the solutions of (\ref{Hom-pi-S_d-Klein}) over the order
of the permutation group:
\be
H_{2-\textsl{g}^*}(\Delta^1,\dots,\Delta^k)=\# \frac{1}{d!}
\{R_1,\dots,R_{\textsl{g}^*},A_1,\dots,A_k \in S_d| A_i\in \textsc{C}_{\Delta^i}  \}
\ee

For instance, take $k=1$, $\Delta^1=(1^3)$ and $\textsl{g}^*=1$ (the number $2-\textsl{g}^*=1$ is the Euler characteristic
of the real projective plane $\mathbb{RP}^2$ and the number $H_1((1^d))$ counts $d$-sheeted unbranched covers
of $\mathbb{RP}^2$, see the Appendix \ref{Counting-of-branched-covers} devoted to the 
geometrical definition of Hurwitz numbers).
Then the number of solutions of the equation $R_1^2=1$ in $S_d,\,d=3$ is equal to 4 (three transpositions
and the identity element).
$H_{1}((1^3))=\frac{4}{3!}=\frac{2}{3}$ (compare to Example \ref{RP^2=no-crit-point} 
in Appendix \ref{Counting-of-branched-covers}).

$$
\,
$$

\paragraph{Mednykh formula \label{Mednykh formula}}
It was found in the papers of A.Mednykh \cite{M1}, Mednykh and Pozdnyakova \cite{M2} (and also in 
\cite{GARETH.A.JONES})
that in both orientable and non-orientable cases there is the unique formula for Hurwitz numbers
in tems of characters of the symmetric group. It depends on the Euler characteristic of the base 
surface $\textsc{e}^*$ and the set of ramification profiles
$\Delta^i$ in critical points as follows:
\be\label{MednykhHurwitz}
H_{\textsc{e}^*}(\Delta^1,\dots,\Delta^k) =\sum_{\lambda} 
\left(  \frac{{ \di}\lambda}{d!} \right)^{\textsc{e}^*}\,\varphi_\lambda(\Delta^1)\cdots
 \varphi_\lambda(\Delta^k)
\ee
Here $\varphi_\lambda(\Delta^i)=|\textsc{C}_{\Delta^i}|\frac{\chi_\lambda(\Delta^i)}{\di\lambda}$ where
$\chi_\lambda(\Delta^i)$ is the character of the irreducable representation of $S_d$ labelled by
the partition $\lambda$ and evaluated at the cycle class labelled by the partition $\Delta^i$,
${\di}\lambda = \chi_\lambda(1^d)$ is the dimension of this representation and $|\textsc{C}_{\Delta^i}|$ 
is the cardinality of the cycle class $\Delta^i$.

At last, let us introduce the following notation
\be\label{H-E-E^*}
H^{\textsc{e}}_{\textsc{e}^*}\left(\lambda^1,\dots,\lambda^m; k+m\right)=
\sum'_{\Delta^1,\dots,\Delta^k}H_{\textsc{e}^*}(\lambda^1,\dots,\lambda^m,\Delta^1,\dots
,\Delta^k)
\ee
where the summation range is constrained by the Riemann-Hurwitz condition (\ref{RHur}):
$d(\textsc{e}^*-k-m)+
\sum_{i=1}^m\ell(\lambda^i)+\sum_{i=1}^k\ell(\Delta^k)=E $
which denotes the sum of all Hurwitz numbers that enumerate $d$-sheeted covers of the Euler characteristic
$\textsc{e}$ with at most $k$ branch points on the base surface with Euler characteristic $\textsc{e}^*$, and
one ramification profile is fixed as $\lambda$.

\subsection{Network of chord diagrams and its genus \label{Chord-Diagram}}

There a number of stidies of the so-called chord diagrams, for some review see \cite{ZL}.
I will present this topic in a way that is convenient for our purposes.

Consider $ F $ circles (loops), each of which is divided into an even number of clockwise directed 
arcs of alternating color: black and white.
The arcs can be drawn with arrows, respectively black or white. 
In the future (in the Section \ref{Product of complex...}) we will associate
black arcs with matrices from the Ginibre ensemble (alternatively: from a circular ensemble in Subsection
\ref{Product of unitary...}), and white
arcs with source matrices (free parameters of our model). The total number of black (white) arcs is a given 
fixed number $ 2n $.
Note that more often than black arcs, the edges of a polygon are considered naturally
they are separated from each other by vertices instead of white arrows.
(All figures in the form of circles (loops) and polygons, we consider up to homeomorphisms thus, do not distinguish
polygons and circles with arcs).

Each black arc has a single partner among the other black arcs that can belong to either the same or 
different loops. We associate these partners with the lines. In the Section \ref{Product of complex...}, 
these partners will be hermitian conjugate matrices, and the lines indicate the pairing in the statistical ensemble.
We call the lines connecting arcs belonging to one loop, {\em chords} and lines connecting arcs belonging to different 
loops, {\em links}.

A connected set of the loops discribed above together with chords and links we will call a network chord diagram 
or simply a {\em network} for the sake of brevity.

Let us describe the procedure which may be called "cutting and joining" loops of the network 
by contracting chords and links:
\begin{itemize}
 \item 
We contract a chord and get two loops, where we preserve the order of the arrows
 \item
We contract a link and naturally unify two loops into one, also preserving the order of the arrows. 
\end{itemize}
Let us remove in $n$ steps all the links and chords in any order. In the end, we get a set consisting of $ V $
loops without chords, which are not connected by links. The number of these loops does not depend on
of the order in which we carry out these actions, see below the Lemma \ref{Lemma1}.

We denote $\tilde{g}^*$ the number of links which we contract
along the cutting-and-joining procedure.
Let us note that we get the following relation
\be\label{F-n-tilde-g*}
V=F+n-2\tilde{g}^*
\ee
Indeed, in the begining we have $F$ loops. Each cutting action adds one loop and each joinning action
removes one loop.

Next, let us introduce the number $\textsc{e}^*:=F-n+V$ and the number $g^*$ related to $\textsc{e}^*$
via $\textsc{e}^*=2-2g^*$. We get $\textsc{e}^*=2F-2\tilde{g}^*$ and $g^*=F-1+\tilde{g}^*$.

The meaning of $\textsc{e}^*$ and of $g^*$ will be clear from the following consideration:

We describe this process in more detail from a different point of view (as the creation of the so-called
ribbon graph (also known as the fatgraph and the embedded graph)):

\begin{itemize}
 \item 
When we contract a chord, 
we attract together  two black chord-partners
(glue together with rubber glue between) so that the beginning of one arrow corresponds to the end of 
the arrow-partner. One can see it as the strip bounded by oppositely directed arrows which becomes  the 
("rubber made")  first edge of the ribbon graph (the same: of the embedded graph, of the fatgraph). 
Thus, we divide the loop into two ones, 
keeping the order of all the remaining arrows. Note that in each loop obtained, we get more white arcs in 
comparison with blach ones.  Notice that we do not tear the chain of arrows-arcs of the initial loop 
and can make a roundtrip following the arrows in its original order and the part of this roundtrip belong to the 
boundary of the new edge (the edge of future ribbon graph). 
The interior of the initial loop turn into 
the interiors of the new loops and of the new fat edge.

\item When we contract a link we glue arcs-partners that belong to different loops again in the way
that the beginning of one arrow corresponds to the end of the arrow-partner. In this case we also get
an edge of the ribbon graph as a strip bounded by to oppositely durected black arrows.

\item Finally, we glue all pairs the black arcs and get the so-called embedded (ribbon) graph 
(see, for example, \cite{ZL})
the vertices of which in our case consist of loops (or, if you like, polygons). This graph consists of strips and 
vertices and can be placed on a Riemann surface (for instance, see \cite{ZL}). One calls the genus $g^*$ and 
the Euler characteristic $\textsc{e}^*$ to the original system of loops, chords and links  (and also the genus and 
the Euler characteristic of the ribbon graph) genus and the Euler characteristic of this Riemann surface: 
${g}^*(\Gamma)={g}^*(\tilde{\Gamma})$, $\textsc{e}^*(\Gamma)=\textsc{e}^*(\tilde{\Gamma})$
It is defined as $ \textsc{e}^* = V-n + F $, where $ V $ is the number of vertices, $ n $ is the number of edges and
$ F $ is the number of faces (domains homeomorphic to a disk and bounded by edges of a graph). The ribbon graph
performs a "triangulation" of the Riemann surface.

\item More about the vertices: if we forget about the edges of the ribbon graph, we'll see a system of 
$ V $ loops, each of which consists of white arcs (arrows pointing clockwise). If we regard it as a polygon, 
replacing the arcs with edges, then from each vertex of such a polygon, the edges of the ribbon graph are 
emitted. Each arrow of the white loop follows the black arrow of the border of the strip emerging from the 
vertex of the polygon that preserves the original order of the arrows. Therefore, "chord diagrams without chords", 
mentioned above, as an end result of the cutting and joining procedure should be considered as the vertices of the 
ribbon graphs

\item Let us number the pairs of white arrows that directly follow the black arrow-partners and assign symbols
$ C_i, C_i ^ * $ for each pair, $ i = 1, \dots, n $. Let's go around a given loop-vertex and enter the {\em word} 
attached to this vertex that we will compose as the product of the symbols from the set 
$ \{C_i, C_i ^ *, \, i = 1, \dots, n \} $ in the order
in accordance with the order of the white arrows on the loop-vertex (we define the product of the symbols up to
cyclic permutations). We get $ V $ words attached to the vertices $ V $ of the ribbon graph. The length of 
each word is equal to the number of edges of the ribbon  graph going from this vertex

\end{itemize}

Thus, cutting-and-joining procedure results in the creation of the ribbon  graph from
a network chord diagram.
If we denote the network $\Gamma$ and the ribbon graph $\tilde{\Gamma}$ then the cutting-and-join procedure
may be symbolically written as
\[
 \Gamma\quad \to \quad \tilde{\Gamma}
\]
The network may be characterized by the data $D_\Gamma$ which are the number of faces $F$, the number of
edges $n$, the number of vertices $V$ of the ribbon graph and also the set of words $\tilde{C}_i,\dots,
\tilde{C}_V$.

What we get.
When we approach a given vertex
following the boundary of the edge of the ribbon graph along a chosen black arrow, we encounter a white arrow on
boundary of the loop-vertex. We follow it and move to another edge of the ribbon graph, which is
  black arrow that followed the white on the original loop. Following this black arrow, we move on to
the next white arrow, which is the boundary of another loop-vertex. So we can have a round trip
according to the chosen intial loop. As we see, indeed, the number of faces of the ribbon graph is equal to 
the number of initial loops and the number of edges is the number of pairs of black arrows. Then the Euler
characteristic of the ribbon graph is completely defined by the number of it's vertices.

\begin{Lemma} \label{Lemma1}

\begin{itemize}
 \item There exists $n!$ way to contract all chord and links. The number $\tilde{g}^*$ does not depend on the
 way
 
 \item The number of the vertices of the ribbon graph is equal to $V=F+n-2\tilde{g}^*$. Therefore, the 
 Euler characteristic $F-n+V=:2-2g^*$ of the network chord diagram is equal to $2F-2\tilde{g}^*$ (and the genus $g^*$ is
 equal to $F-1+\tilde{g}^*$).
\end{itemize}

\end{Lemma}

Thus, for a network which consists of a single loop
 with chords $\tilde{g}^*=g^*$.

The first item will be proven at the end of Section \ref{Product of complex...}. 

The proof of the second item is as follows. First, let us make a 

\begin{Remark}
 One can transform a given network 
 to a {\em minor} network by replacing 1) a given neiboring white--black--white arrows by a single white arrow
 2) doing the same with the black partner of the chosen black above arrows. This is the procedure of forgetting
 of a black pair. Then, one can recollect it and insert the pair back.

 One chose the order to perform the creation of the ribbon graph by numbering of black pairs.
 Gluing the first pair he forgets about all other black arrows replacing all of them as explained above.
 He gets one edge and two white arrows which form either a single, or two white loop-vertices.
 This is the simplest ribbon graph.
 Then he  recollect the second black pair and gets the second ribbon graph. Thus one gets the sequence
 of ribbon graphs defined by the numeration of the steps.
 
\end{Remark}

\be\label{straight-chain}
\Gamma \,\to\,\Gamma_1 \,\to\,\Gamma_{1,2}\,\to\cdots\,\to\,\Gamma_{1,\dots,n}=\tilde{\Gamma}
\ee
One can chose another consequence of steps which is obtaines by the re-enumaration of $1,\dots,,n \,\to
\sigma(1),\dots,\sigma(n)$, $\sigma\in S_n$:
\be\label{other-chains}
\Gamma \,\to\,\Gamma_{\sigma(1)} \,\to\,\Gamma_{\sigma(1),\sigma(2)}\,
\to\cdots\,\to\,\Gamma_{\sigma(1),\dots,\sigma(n)}=\tilde{\Gamma}
\ee
There exists $n!$ paths to achieve $\tilde{\Gamma}$ and there are $\frac{n!}{k!(n-k)!}$ different
$\Gamma$ with $k$ subscripts.

Having this remark in mind we see that each cutting step (contraction of the chord) results in adding of 1 
edge to the ribbon graph and also of 1 loop-vertex. While each joining step (contraction of the link) 
results in adding of 1 edge and removing of 1 vertex. We have $F$  vertices in the beginning and $n$ steps to create the final
ribbon graph. Therefore, at the end we get $V=F+n-\tilde{g}^*$ vertices.

This Lemma together with Lemma \ref{useful-relations} is important.

As is well known after the papers of Kazakov, Bresin \cite{BrezinKazakov}, Migdal and Gross
\cite{MigdalGross} (see \cite{ZL} for a review
which emphasizes mathematical aspects), the ribbon graphs can be listed using
models of Hermitian matrices. In our case (see Section \ref{Product of complex...})
the ribbon graph will initially be specified by the choice of the matrix model.
Thus, for each Feynman graph of the one-matrix model we assigh the matrix model labeled
by this graph.

\subsection{Random matrices. Complex Ginibre ensemble}

\paragraph{Complex Ginibre ensembles.}
On this subject there is an extensive literature, for instance see \cite{Ak1,Ak2,AkStrahov,S1,S2}.

We will consider integrals over $N\times N$ complex matrices $Z_1,\dots,Z_n$ where the measure is defined as
\be\label{CGEns-measure}
d\Omega(Z_1,\dots,Z_n)= \prod_{\alpha=1}^n d\mu(Z_\alpha)=c_N^n 
\prod_{\alpha=1}^n\prod_{i,j=1}^N d\Re (Z_\alpha)_{ij}d\Im (Z_\alpha)_{ij}\text{e}^{-N|(Z_\alpha)_{ij}|^2}
\ee
where the integration range is $\mathbb{C}^{N^2}\times \cdots \times\mathbb{C}^{N^2}$ and where $c_N^n$
is the normalization 
constant defined via $\int d \Omega(Z_1,\dots,Z_n)=1$.

We treat this measure as the probability measure. The related ensemble is called the ensemble 
of $n$ independent complex Ginibre enesembles. 
The expectation of a quantity
$f$ which depends on entries of the matrices $Z_1,\dots,Z_n$ is defined by
$$
\mathbb{E}_{n,N}(f)=\int f(Z_1,\dots,Z_n) d\Omega(Z_1,\dots,Z_n).
$$
The subscript $ n $ reminds that the expectation is estimated in the product of $ n $ independent
Ginibre ensembles, and the second subscript, $ N $, - that the Gauss measure is not chosen as
$ e^{- \tr ZZ^\dag} $, but in the form $ e^{-N \tr ZZ^\dag} $.

\paragraph{Spectral correlation functions.} For any given matrix $X$ and a partition 
$\lambda=(\lambda_1,\lambda_2,\dots,\lambda_\ell)$ we introduce the following
notations
\be\label{p(X)}
{\bf p}(X)=\left(\tr X , \tr X^2 , \tr X^3,\dots \right)
\ee
\be\label{p-lambda-(X)}
{\bf {p}}_\lambda (X)=\tr X^{\lambda_1}  \tr X^{\lambda_2} \cdots \tr X^{\lambda_\ell}
\ee
Each $\tr X^{\lambda_i}$ is the Newton sum $\sum_{a=1}^N x_a^{\lambda_i}$ of the eigenvalues 
$x_a,\,a=1,\dots,N$ of the matrix $X$. 

We are interested in the spectral correlation functions
$\mathbb{E}_{n,N}({\bf p}_{\lambda^1}(X_1)\cdots {\bf p}_{\lambda^m}(X_m))$ where $X_i,\,i=1,\dots,m$
is a set of matrices and 
$\lambda^i=(\lambda^i_1,\lambda^i_2,\dots),\,i=1,\dots,m$ 
is a set of given partitions.\footnote{Throughout the paper the upper 
index of the parts partitions is not a power but a label which
indexes different partitions.}

Let us introduce the notations ${\bf p}=(p_1,p_2,\dots)$ which is the 
semiinfinite set of parameters and
\be
\texttt{V}(X,{\bf p})=\sum_{n>0} \frac 1n p_n X^n 
\ee
Then it is well-known that
\be\label{Taylor-for-e^V}
e^{N\tr \texttt{V}(X,{\bf p})} = \sum_\Delta \frac{1}{z_\Delta} N^{\ell(\Delta)} {\bf p}_\Delta(X){\bf p}_\Delta
\ee
where the sum ranges over all partitions $\Delta=(\delta_1,\delta_2,\dots, \delta_k)$, $\delta_k >0$,
$k=0,1,3,\dots$ and 
$\ell(\Delta)$ denotes the {\it length} of the partition $\Delta$, i.e. the number of the non-vanishing
parts of $\Delta$.
. The notations are
as follows: ${\bf p}_\Delta=p_{\delta_1}p_{\delta_2}\cdots$, and 
$z_\Delta = \prod_{i=1}^\infty i^{m_i}m_i!$ where $m_i$ is the number of parts
$i$ which occur in the partition $\Delta$. For instance, for the partition $\Delta=(5,5,2,1,1)$
we get $z_\Delta=5^2 \times 2!\times 2 \times 1!\times 1^2\times 2!  = 200$.

\begin{Remark} \label{Taylor-for-prod-e^V}
Let us note that the generation function of the spectral invariants may be choosen as
\be
\mathbb{E}\left( e^{N\tr \texttt{V}(X_1,{\bf p}^{(1)})}\cdots e^{ N \tr \texttt{V}(X_m,{\bf p}^{(m)})}  \right)
\ee
Indeed, with the help of (\ref{Taylor-for-e^V})  the Taylor series in parameters ${p}_k^{(i)}$ 
yields the mentioned spectral correlation functions.
\end{Remark}

\subsection{Hypergeometric tau functions}

\paragraph{Schur functions.}
In what follows we need polynomials in many variables
called functions of Schur labeled by partitions \cite{Mac}. First, we introduce the so-called elementary Schur functions
$ s_{(n)} $, labeled by partitions $ (n) $ with one part equal to $ \lambda_1=n $, which are defined as follows:
$$
e^{\texttt{V}(x,{\bf p})}=\sum_{n\ge 0} x^n s_{(n)}({\bf p})
$$
In particular, $s_{(0)}({\bf p})=1$,  $s_{(1)}({\bf p})=p_1$, $s_{(2)}({\bf p})=\frac12 (p_1^2+ p_2)$.

Schur function $s_\lambda$ labeled by a given partition $\lambda=(\lambda_1,\dots ,\lambda_N)$ is defined
in terms of the elementary ones by
\be\label{SchurFunction}
s_\lambda({\bf p}) = \det \left( s_{(\lambda_i-i+j)}({\bf p}) \right)_{i,j}
\ee

We shall write the Schur function also as the function of matrix argument which we write as a capital letter
 say $X$ having in mind that it is $s_\lambda(X):=s_\lambda({\bf p}(X))$ where 
 ${\bf p}(X)=\left(p_1(X),p_2(X),\dots  \right)$ with $p_n(X)=\tr X^n$. If $x_1,\dots,x_N$ are
 the eigenvalues of the $N\times N$ matrix $X$ then $s_\lambda(X)$ is the symmetric homogenious polynomial 
 in eigenvalues and can be written as
 \be\label{Schur-in-X}
 s_\lambda(X)= \frac{\det \left(x_j^{N+\lambda_i-i}  \right)}
 {\det \left(x_j^{N-i} \right)}
 \ee
 The formula known as Cauchi-Littlewood relation is very useful
 \be\label{C-L}
 e^{N \tr \texttt{V}(X,{\bf p})} =\sum_\lambda s_\lambda(X)s_\lambda(N{\bf p})
 \ee
 where the sum ranges over all partitions whose length (the number of non-vanishing parts) does not exceed $N$,
 and $N{\bf p}:=\left( Np_1, Np_2, Np_3,\dots \right)$.

\paragraph{Degree and Euler characteristic}. 
For each ratio of Schur functions labeled with the same
partition, we assign the {\em degree} $ {\deg} $ as follows
\be\label{degree}
{\deg}
\left( 
\prod_i \left( s_\lambda(A_i) \right)^{{\rm d}_i}
\right)=\sum_i {\rm d}_i
\ee 
 As follows from the Mednykh formulas (\ref{Hom-pi-S_d-Riemann}) and (\ref{Hom-pi-S_d-Klein}),
sums over all $ \lambda $ of such expressions can be used to generate the Hurwitz numbers, where
the degree gives the Euler characteristic of the base surface.
 
\paragraph{Content product.} For a given number $x$ and a given Young diagram $\lambda$ the content 
product is defined as the product
\be\label{Poch}
(x)_\lambda :=\prod_{(i,j)\in \lambda} (x+j-i) 
\ee
The number $j-i$, which is the distance of the node with coordinates $(i,j)$ to the main diagonal of the Young
diagram $\lambda$ is called the {\em content} of the node. For one-row $\lambda$, the content product is the 
Pochhammer symbol $(a)_{\lambda_1}$. For a given function of one variable $r$, we define the generalized
content product (the generalized Pochhammer symbol) as
\be\label{content-product}
r_\lambda(x)= \prod_{(i,j)\in\lambda} r(x+j-i)
\ee
The content product plays an important role in the representation theory of the symmetric groups.
It was used in \cite{OS-2000} to define certain family of tau functions which we called hypergeometric
tau functions.
 
{\bf Example}. The example of the content product may be constructed purely in terms of the Schur functions:
if we choose 
\be\label{Example:q,t}
r(x)=\prod_{i} \left(\frac{1-\texttt{q}_i\texttt{t}_i^x}{1-  \texttt{t}_i^x}  \right)^{\texttt{d}_i}
\ee
where $ \texttt{q}_i,\texttt{t}_i,\texttt{d}_i$ are parameters, we obtain
\be\label{r-rational}
r_\lambda(x)=\prod_i 
\left( \frac{s_\lambda({\bf p} (\texttt{q}_i,\texttt{t}_i))} {s_\lambda({\bf p} (0,\texttt{t}_i))} 
\right)^{\texttt{d}_i}
\ee
One can degenerate (\ref{Example:q,t}) to the rational function and obtain
\be\label{r-lambda-rational}
r_\lambda(x)= \frac{\prod_{i=1}^p ({\texttt{a}}_i)_\lambda}{\prod_{i=1}^q (\texttt{b}_i)_\lambda}=
\prod_{i=1}^p \frac{s_\lambda({\bf p}(\texttt{a}_i))}{s_\lambda({\bf p}_\infty)}
\prod_{i=1}^q \frac{s_\lambda({\bf p}_\infty)}{s_\lambda({\bf p}(\texttt{b}_i))}
\ee
Above we used the following special notations:
\be\label{special-p}
{\bf p}_\infty=(1,0,0,\dots),\quad {\bf p}(\texttt{a})=(\texttt{a},\texttt{a},\texttt{a},\dots),\quad 
p_m(\texttt{q},\texttt{t})=\frac{1-\texttt{q}^m}{1-\texttt{t}^m}
\ee
Actually, any reasonable content product can be interpolated by expressions (\ref{r-lambda-rational}).
Because of this, the degree of content products always vanishes \footnote{In what follows, this leads to 
the fact that in generation functions the content
products do not affect the Euler characteristic  of the base surface, but only affect ramification
type of the covering map.}.

\paragraph{Hypergeometric tau functions of the Toda lattice and two-component KP hierarchy.}
The function
\be\label{tau--hyp}
\tau_r(x,{\bf p},{\bf p}^*):=\sum_{\lambda} r_\lambda(x) s_\lambda({\bf p}) s_\lambda({\bf p}^*)
\ee
solves an infinite number of compatible equations of differential equations, separetely, in
the variables ${\bf p}$ (KP hierarchy), separetely in variables ${\bf p}^*$ (second KP hierarchy) and
also in the variable $x$ which is supposed to be a discrete variable. It was introduced and analized
in details in \cite{OS-2000},
but, in fact, it appeared earlier in \cite{KMMM} in a different way without the usage of content product.
This family of tau functions has numerous applications, some of them are mentions in the Appendices to
to \cite{OS-2000} and to \cite{O-TMP-2006}. The well-know hypergeometric functions in one variable 
(the Gauss one, basic ones, the so-called generalized ones) 
together with certain hypergeometric functions of matrix argument (for instance 
Milne's hypergeometric function) are examples of (\ref{tau--hyp}). 

We will write also $\tau_r(x,{\bf p},X)$ having in mind that 
the Schur function in (\ref{tau--hyp}) is written as a matrix. For instance, if we select the content
product as in example (\ref{r-lambda-rational}), and if we choose the matrix
$X$ to be $1\times 1$ matrix and ${\bf p}$ to be $(1,0,0,\dots)$, we obtain the so-called generalized
hypergeometric function ${_{p}F}_q(\{\texttt{a}_i\},\{\texttt{b}_i\},X)$.

Let us note that we can write the argument of the tau function not as ${\bf p}=(p_1,p_2,\dots$ but
as $N{\bf p}=(Np_1, Np_2, Np_3,\dots)$. In this case the variables $Np_i,\,i>0$ play the role of the higher times
\cite{JM}. This replacement turns out to be suitable in $N\to\infty$ limit. It was used, say, in \cite{NZ} in the study
of Hurwitz numbers generated by the model of normal matrices. It is also suitable for us in view of the 
choice of Gauss measure in the Ginibre ensembles in form presented by (\ref{CGEns-measure}).

The simplest (and the main for our purposes) example is the case $r$ identically is equal to one.
Such tau function will be denoted $\tau_1$. It does not depend on $x$:
\be\label{tau1--hyp}
\tau_1(x,X, N {\bf p})= e^{N\tr \texttt{V}(X,{\bf p})} = e^{N\sum_{i=1}^N \sum_{m>0} \frac 1m x_i^m p_m }=
\sum_\lambda s_\lambda(X)s_\lambda( N {\bf p})
\ee
where $x_1,\dots,x_N$ are eigenvalues of $X$,
in addition, for such tau function we have  (\ref{Taylor-for-e^V}). 

\begin{Remark} \label{Remark-Miwa-case} Let us specify the set of variables 
${\bf p}=(p_1,p_2,\dots)$ in formula (\ref{C-L}) as follows:
 \be\label{Miwa-case}
  p_m=p_m(\{ \texttt{d}_i,x_i  \}):  =-\sum_{i=1}^L \texttt{d}_i x_i^m
  \ee
  Then,
  \be\label{Miwa-det}
  e^{ N \tr \texttt{V}(X,{\bf p})}=\prod_{i=1} \det\left(1- x_iX \right)^{N\texttt{d}_i}
  \ee
  If all $N \texttt{d}_i$ are natural numbers, (\ref{Miwa-det}) is a polynomial function of entries of $X$;
  the right hand sides of (\ref{C-L}) and of (\ref{Taylor-for-e^V}) have a finite number of terms. 
  In this case, as follows from the properties Schur functions, see (\ref{p-to-p-in-Schur}) in  
  Appendix \ref{Partitions-and-Schur-functions}
$ s_\lambda(N{\bf p}) = 0 $ if $ \lambda_1> N\sum_i \texttt{d} _i $ and tau function 
(\ref{tau--hyp}) is also a polynomial.

 \end{Remark}

\paragraph{Hypergeometric tau function of the BKP hierarchy.}
The expression
\be\label{tau--hypB}
\tau_r^{\rm B}(M,x,{\bf p}):=
\sum_{\lambda\atop \ell(\lambda)\le M} r_\lambda(x) s_\lambda({\bf p}) 
\ee
is also a tau function but now it 
is a tau function of the hierarchy introduced in \cite{KvdLbispec},
which authors called the "fermionic" BKP hierarchy and we call the "large" BKP hierarchy (to make difference
with the BKP hierarchy invented in \cite{JM}).
Tau function (\ref{tau--hypB}) appeared in \cite{OST-I}.
The simplest (and most impostant for us) example
is again the case where $r$ is identically equal to 1 and $M=\infty$:
\be\label{tau1--hypB}
\tau_1(X)=\sum_\lambda s_\lambda(X) =
\ee
\[
 e^{ \sum_{m>0} \frac {1}{2m} \left(\tr X \right)^{2m} 
+ \sum_{m\,{\rm odd}}\frac1m\tr X^{m}}=\prod_{i=1}^N (1-x_i)^{-1}\prod_{i<j}^N (1-x_ix_j)^{-1}
\]
where $x_1,\dots,x_N$ are eigenvalues of $X$.
 
Notice that 
\be\label{deg-tau}
\deg\left( \tau_r\right)=2,\quad \deg \left( \tau_r^{\rm B}  \right)=1
\ee

\section{Products of complex and random matrices and certain sums  related to chord diagrams and Hurwitz numbers
\label{Product of complex...}}

The expression
\[
 Z_1Z_2 \cdots Z_{n-1}Z_nZ_n^\dag Z_{n-1}^\dag \cdots Z_2^\dag Z_1^\dag
\]
where random matrices $Z_i,\, i=1,\dots,n$ belong to $n$ independent complex Ginibre ensembles
was the object of study in numerous papers (in particular, in relation to quantum chaos and to transmition problems
see \cite{Ak1},\cite{Ak2},\cite{AkStrahov},\cite{S1},\cite{S2},\cite{Alfano}, in relation to
Hurwitz numbers see \cite{Chekhov-2014}, \cite{NO-2014} in relation to tau functions see
\cite{OrlovStrahov}).

We want to consider modifications of this product, namely, let us: 

\begin{itemize}
 \item add constant (the ''source'') matrices
between random ones: $Z_i\to Z_iC_i$, $Z_i^\dag \to Z_i^\dag C_i^*$
  \item permute the order in the product in an arbitrary way which we encode
by a chord diagram
  \item  factorize this product into $F$ factors and introduce network
chord diagram to encode it

\end{itemize}

\subsection{The model of complex matrices labeled by a network}

Consider a set of $N\times N$ matrices $\{Z_1 C_1,Z_1^\dag C_1^*,Z_2 C_2,Z_2^\dag C_2^*,\dots,Z_n^\dag C_n^*\}$, where
$C_1,\dots,C_n, C_1^*,\dots,C_n^*$ are given complex matrices (source matrices) and  each 
of $Z_i$, $ i=1,\dots,n$ belongs to the $i$-th complex Ginibre ensemble.
Here and below, the dag denotes Hermitian conjugation, and $C_i^*$ is unrelated to $C_i$.
Notice that each matrix from Ginibre ensemble is multiplied from the right by the source matrix
with the same number. The order in the
Thus, we consider a product of $2n$ matrices where the matrices $Z_iC_i$
and $Z_i^\dag C_i^*$ enter in a given order. Each of the written above $2n$ matrices enters the product 
only {\em once}, and this condition is important in what follows. 
We denote this product $X$. Each possible product $X$ can be presented graphically as a loop
with $2n$ black directed arcs and $2n$ white directed arcs as we explained in Subsection \ref{Chord-Diagram},
black arrows are related to random matrices and white arrows are related to the source matrices.
Each pair of hermitian conjugate random matrices is associated by the chord. This is the case of the single
loop (the chord diagram), that is $F=1$ as it explained in Subsection \ref{Chord-Diagram}. The general 
case related to the network of chord diagrams is obtained by splitting  this product into factors (sub-products)
$X=X_1\cdots X_F$ in a way that the source matrices are nearest right neibours of each
$Z_i$ and to each $Z_i^\dag$ as it was before, and we also ask the obtained network to be connected. 

Thus, we have a given network, say $\Gamma$, which defines matrix 
products in $X_1,\dots,X_F$ and related ribbon graph $\tilde{\Gamma}$ equipped with data $D_\Gamma$,
namely, the number of faces $F$, the number of edges $n$, the number of vertices $V$ (and the Euler
characteristic $\textsc{e}^*$ equal to $F-n+V$) and the set of words
$\tilde{C}_1,\dots,\tilde{C}_V$. Then

\begin{Theorem}\label{Theorem-1} 
Consider the set of tau functions (\ref{tau--hyp}):
$$\tau_{r^{(1)}}\left(x,X_1, N {\bf p}^{(1)}\right),\,\dots,\,\tau_{r^{(F)}}\left(x,X_F, N {\bf p}^{(F)}\right)$$
defined by the set of given functions $r^{(1)},\dots,r^{(F)}$, which depend on the matrix products $X_1,\dots,X_F$
described by the network $\Gamma$ with data $F,n,V$ and $\tilde{C_1},\dots,\tilde{C}_V$.
Consider the expectation value of the product of these tau functions in $n$ independent Ginibre ensembles.
 \be\label{Th-1}
\mathbb{E}_{n,N}\left( \prod_{a=1}^{F} \tau_{r^{(a)}}(x,X_a,N{\bf p}^{(a)}))
\right) =
\ee
\be\label{Th-1-in-Schur-series}
\sum_{\lambda\atop \ell(\lambda)\le N} 
r_\lambda(x) \left(  
s_\lambda(N{\bf p}_\infty) 
\right)^{-n} \prod_{a=1}^{F} s_\lambda(N{\bf p}^{(a)})
\prod_{a=1}^{V}  s_\lambda({\tilde C}_a) 
 \ee
 where $r_\lambda(x)$ is the content product (\ref{content-product}) where  $r=\prod_{a=1}^{F} r^{(a)}$,
 where ${\bf p}_\infty:=(1,0,0,\dots)$,
 and ${\bf p}^{(a)}=(p^{(a)}_1,p^{(a)}_2,\dots ),\, a=1,\dots,F$ are sets of parameters. 
 
 The degree of (\ref{Th-1-in-Schur-series}) $F-n+V$ coinsides with Euler characterisitic of the network 
 $\textsc{e}^*(\Gamma)$.
\end{Theorem}
\begin{Remark} Remark 1.
Note that for general values of the parameters $ {\bf p}^{(a)} $ both (\ref{Th-1}) and 
(\ref{Th-1-in-Schur-series}) diverge. However, there are open domains of these variables (parameterized by numbers
$ NL, N\texttt{d}_1, \dots, N\texttt{d}_L \in \mathbb{N} $ and $ x_1, \dots, x_L \in \mathbb{C} $ from
(\ref{Miwa-case})), where both (\ref{Th-1}) and (\ref{Th-1-in-Schur-series}) are finite.
\end{Remark}

Notice that if we choose the function $r$ to be in form (\ref{Example:q,t}) the series (\ref{Th-1-in-Schur-series})
is written only in terms of the Schur functions. (\ref{Th-1-in-Schur-series}) generalizes the Hurwitz generating 
series suggested in \cite{AMMN-2014}.

In certain cases the integral of tau functions (\ref{Th-1}) is a tau function itself, 
 however these cases are related to $\textsc{e}^*=2,1$ see for instance, Examples 2,3,4 in this Subsection. 

\begin{Remark} Remark 2.
 Notice that the degree of the product of the tau functions in (\ref{Th-1}) is equal to $2F$,
 while the degree of (\ref{Th-1-in-Schur-series}) is $F-n+V=:2F-2\tilde{g}^*$ where $\tilde{g}^*\ge 0$.  
\end{Remark}

\begin{Corollary}\label{Prop1} For $F=1$ and $r=1$ case, we get
\be\label{Pr1}
\mathbb{E}_{n,N} \left( e^{N\tr \texttt{V}(X,{\bf p})}\right)=
\sum_{\lambda}\left(  
s_\lambda(N{\bf p}_\infty) 
\right)^{-n} s_\lambda(N{\bf p})
\prod_{i=1}^{V}  s_\lambda({\tilde C}_i)
\ee
In particular, if all sources are $N\times N$ identity matrices we
get
\be\label{Pr1-I_N}
\mathbb{E}_{n,N} \left( e^{N\tr \texttt{V}(X,{\bf p})}\right)=
\sum_{\lambda}\left(  
s_\lambda(N{\bf p}_\infty) 
\right)^{-n} s_\lambda(N{\bf p})
  \left(s_\lambda(\mathbb{I}_N)\right)^{V}
\ee
where $s_\lambda(\mathbb{I}_N)=(N)_\lambda s_\lambda({\bf p}_\infty) $, for the notation $(N)_\lambda$ see (\ref{Poch}).
\end{Corollary}

{\bf Example 1}. Consider the product $X=Z_1C_1Z_2C_2Z_1^\dag C_1^* Z_2^\dag C_2^*  $ which is related to the chord
diagram with two intersecting chords. As we can find in this case $F=1,\, n=2,\, V=1$ (so, $\textsc{e}^*=0$ which
is related to the torus) and we get a single word
equal to $C_2C_1C_2^*C_1^*$. (Thus, we get 4 edges of the ribbon graph coming from the single vertex). 
In case all source matrices were $\mathbb{I}_N$, we obtain 
$\sum_\lambda (N)_\lambda s_\lambda(N{\bf p}) \left(s_\lambda(N{\bf p}_\infty)\right)^{-1}   $ in the right hand side
of (\ref{Pr1-I_N}).

{\bf Example 2} Consider $X=Z_1C_1Z_2 C_2\cdots Z_nC_n  (Z_n^\dag C_n^* \cdots Z_2^\dag C_2^* Z_1^\dag C_{1}^*  )$
\footnote{\label{Dong-Wang} In case where $C_1^*, C_n$ are Hermitian and $C_i=C_{i+1}^\dag,\,$
$C^*_i=C^\dag_i,\,i=1,\dots,n-1$ the matrix $X$ is Hermitian and it is the only case of Hermitian $X$.}.
This is an example of a chord diagram where chords do not intersect. It is easy to show that in such case we 
always have $\textsc{e}^*=2$. The set of words consists of $V=n+1$ matrices: 
$C_n,\,C_1^*$ and of $C_iC_{i+1}^*,\,i=1,\dots,n-1$.
The ribbon graph is the linear tree graph. In case all source matrices are identity ones (therefore, $X$ is Hermitian), 
we get  $\sum_\lambda \left( (N)_\lambda \right)^{n+1} s_\lambda(N{\bf p}) s_\lambda(N{\bf p}_\infty)   $
 in the right hand side of (\ref{Pr1-I_N}) that is tau function (\ref{tau--hyp}) where $r(x)=x^{n+1}$, and
 this case was carefully studied, in particular see \cite{Ak1}, \cite{Ak2}, \cite{AkStrahov}.

{\bf Example 3}. Consider $X=(Z_1C_1Z_1^\dag C_1^*)\cdots (Z_nC_nZ_n^\dag C_n^*)$. 
This is another example of chord diagram where chords do not intersect. The number of vertices is equal to $n+1$. 
The words are $C_1,C_2,\dots,C_n$ and $C_1^*C_2^*\cdots C_n^*$ (thus,
 $n$ edges of the ribbon graph come out of the single nontrivial vertex. This is a star-like ribbon graph
 drawn on the Riemann sphere). In case all source matrices are identity ones, 
$X$ is the product of positive Hermitian matrices. In that case 
 we get the same answer
 $\sum_\lambda \left( (N)_\lambda \right)^{n+1} s_\lambda(N{\bf p}) s_\lambda(N{\bf p}_\infty)   $
 in the right hand side of (\ref{Pr1-I_N}) as in the previous Example.

 Other examples of the $F=1$, in particular related to the case $\textsc{e}^*$ may be found in the  
 previous work \cite{O-2017}.

Now consider the case where $r=1$ with sets of faces $F>1$.

\begin{Corollary}\label{Prop2}
 \be\label{Pr2}
\mathbb{E}_{n,N}\left( e^{N \tr \texttt{V}(X_1,{\bf p}^{1})}\cdots e^{N \tr \texttt{V}(X_F,{\bf p}^{F})} \right) =
\sum_{\lambda}\left(  
s_\lambda(N{\bf p}_\infty) 
\right)^{-n} \prod_{i=1}^F s_\lambda(N{\bf p}^{(i)})
\prod_{i=1}^{V}  s_\lambda({\tilde C}_i)
 \ee
In particular, if all source matrices are equal to $\mathbb{I}_N$ we get 
 \be\label{Pr2-I_N}
\mathbb{E}_{n,N}\left( e^{N\tr \texttt{V}(X_1,{\bf p}^{1})}\cdots e^{N\tr \texttt{V}(X_F,{\bf p}^{F})} \right) =
\sum_{\lambda}
\left(s_\lambda(\mathbb{I}_N) \right)^V
\left(s_\lambda(N{\bf p}_\infty) \right)^{-n}  
\prod_{i=1}^F s_\lambda(N{\bf p}^{(i)})
 \ee
 
 \end{Corollary}
 
 Let us notice that if $N=1$, the matrices commute, and the answer does not depend on the order
 in the product that defines the number $V$. And we see that it is the case because each $s_\lambda(\mathbb{I}_N)=1$.
 
Thus, the number of the factors $s_\lambda({\tilde C}_i)$ is the number of vertices, the number of
factors $s_\lambda(N{\bf p}_\infty) $ is the number of edges, and the number of factors
$ s_\lambda(N{\bf p}^{(i)})$ is the number of faces of the ribbon graph. The formula (\ref{Pr2})
is nice. In \cite{OS-TMP} we appreciate the expression of hypergeometric tau functions written
only in terms of the Schur functions, which obtained if we use the content product (\ref{r-lambda-rational}).

 By (\ref{Taylor-for-e^V}) (choosing only $|\lambda|=1$ terms in the right 
hand side of (\ref{Pr2})), we get 
\begin{Corollary} 

\be\label{expect-traces}
\mathbb{E}_{n,N}\left( \tr X_1 \cdots \tr X_F \right) = N^{-n}\prod_{i=1}^{V}  \tr {\tilde C}_i 
\ee
In case all sources are identity $N\times N$ matrices, we obtain 
\be\label{expect-traces-I_N}
\mathbb{E}_{n,N}\left( \tr X_1\cdots \tr X_F  \right) = N^{V-n}=N^{\textsc{e}^*-F}
\ee

\end{Corollary}

It follows from this Corollary then the expectation value in the right hand side of 
(\ref{expect-traces}) grows with $N$ only in case $\textsc{e}^*=2$ (Riemann sphere)
and $F=1$. Otherwise the right hand side of (\ref{expect-traces-I_N}) vanishes if $N\to\infty$.

From this Corollary it follows that the expectation on the right-hand side of
(\ref{expect-traces}) grows together with $ N $ only in the case $ \textsc{e}^* = 2 $ (the Riemann sphere)
and $ F = 1 $.

{\bf Example 4}. Take  $F=2$ and $X_1=Z_1C_1\dots Z_nC_n$, $X_2=Z_n^\dag C_n^* \cdots Z_1^\dag C_1^*$. 
As one can see in this case $V=n$ (thus, $\textsc{e}^*=2$) and the words are $C_iC_i^*,\,i=1,\dots,n$.
We have two faces (regions delimited by the graph).
The right hand side of (\ref{Pr2}) is equal
to 
$$
\sum_\lambda  s_\lambda(N{\bf p}^{(1)}) s_\lambda(N{\bf p}^{(2)}) \prod_{i=1}^{n} 
\frac{s_\lambda(C_iC_{i+1}^*)}{s_\lambda(N{\bf p}_\infty)},\quad C_{n+1}^*:=C_1^*
$$
In case all source matrices were $\mathbb{I}_N$ it is equal to 
$\sum_\lambda \left( (N)_\lambda \right)^{n} s_\lambda(N{\bf p}^{(1)}) s_\lambda(N{\bf p}^{(2)})  $
which is tau function (\ref{tau--hyp}) with $r(x)=x^n$.

{\bf Example 5}. Take $F=n$ and $X_a=Z_aC_aZ_{a+1}^\dag C_{a+1}^*,\,1\le a < n$ and 
$X_F=Z_nC_nZ_1^\dag C_1^*$
(a closed chain). We obtain two vertices (so, $\textsc{e}^*=2$) and two words
$\tilde{C}_1=C_1C_2\cdots C_n$, $\tilde{C}_2=C_n^*C_{n-1}^*\cdots C_1^*$. In case all source matrices
were $\mathbb{I}_N$, we get that the right hand side of (\ref{Pr2}) is equal to
$$\sum_\lambda \left( (N)_\lambda \right)^{2} \left(s_\lambda(N{\bf p}_\infty)\right)^{2-n}
\prod_{a=1}^n s_\lambda(N{\bf p}^{(a)}). $$ It can be identified with the tau function (\ref{tau--hyp}), if we 
fix each set ${\bf p}^{(a)},\,a=1,\dots,n$  to be in the form (\ref{special-p}), with the exception of the selected
two that we will interpret as higher times of the two-component KP hierarchy.

\paragraph{About certain sums.}

Consider the sum
\be\label{Y}
Y=\sum_{i=1}^n \left( Z_iC_i + Z_i^\dag C_i^* \right)
\ee
where matrices $Z_i,\, i=1,\dots, n$ belong to $n$ independent complex Ginibre ensembles, and
complex matrices $C_i,\,i=1,\dots$, plays the role of sources. Let us split the sum $Y$
into the sum of $v$ terms $Y=Y_1+\cdots +Y_F$. Denote $k_i$ the number of terms in $Y_i$,
and denote $J_i$ the collection of all matrices from the set $\{Z_aC_a,Z_a^\dag C_a^*,\,a=1,\dots,n \}$ 
that enter $Y_i$.
We have $k_1+\cdots +k_F=2n$. 
For instance, $Y_1=Z_1C_1+Z_2C_2+Z_n^\dag C^*$ and $Y_2=Y-Y_1$; then, $k_1=3,\,k_2=2n-3$
We denote the subset of matrices which enter $Y_i$ by $J_j$.

Let us rescale $C_i\to a_i^{-1}C_i$. 
 Consider
\be\label{corr-of-sums}
\mathbb{E}_{n,N}\left( \tr Y_1^{m_1} \cdots \tr Y_F^{m_F} \right) = {\rm Pol}(a^{-1})
\ee
where $m_i\le k_i$ and  $m_1+\cdots + m_F\le 2n$. The right hand side of expression (\ref{corr-of-sums})
is written to notify that it is a polynom in $a_1^{-1},\dots,a_n^{-1}$. Monomials which are multilinear 
in $a_1^{-1},\dots,a_n^{-1}$ may be evaluated with the help of relation (\ref{expect-traces}). Indeed, 
thanks to the summation in the right hand side of (\ref{Y}),
the left hand side of (\ref{corr-of-sums})
is the sum of many terms which are monomials bilinear in random matrices $Z_i$ and $Z_i^\dag$. Each monomial 
obtained in this way may be written as $\tr X_1 \tr X_2 \cdots \tr X_F$, where each $X_i$ is a product
of matrices $Z_{i_j}C_{i_j}$ and $Z_{k_j}^\dag C_{k_j}^* $ from the subset of matrices $J_j$ which enter $Y_j$.
To apply (\ref{expect-traces}) one needs the requirement that each of $\{ Z_iC_i,Z_i^\dag,\,i=1,\dots,n \}$ enters
the product $X_1\cdots X_F$ at most once. We get it by picking up residium terms in the right hand side
of (\ref{corr-of-sums}) which is a polynom in $a_i^{-1},\,i=1,\dots,n$.
We obtain
\[
 {\rm res}_{a_1}\cdots {\rm res}_{a_s}\,\mathbb{E}_{n,N}\left( \tr Y_1^{m_1} \cdots \tr Y_F^{m_F} \right)
 = N^{-n} \sum_{\Gamma}' \prod_{i=1}^{V}  \tr {\tilde C}_i(\Gamma) 
\]
where $\Sigma'_\Gamma$ denotes the sum over $k_1!\cdots k_F!$ networks of chord diagrams with $F$ loops
which are obtained by all permutations of endpoints of chord and links along each of loops
(which encode all permutations of the matrices in the sets $J_i$), where diagrams obtained by cyclic
permutation along loops give the same contribution. For the case where all source matrices are identity ones,
in $N\to\infty$ limit the mail contribution proportional
to $N^{2-F}$ give diagrams with $\textsc{e}^*=2$ (see (\ref{expect-traces-I_N})), and the main term is equal to $c_2(F)N^{2-F}k_1\cdots k_F$,
where  $c_{\textsc{e}^*}(F)$ is the number of chord diagrams with $F$ faces and the Euler characteristic 
equal to $\textsc{e}^*$.

\subsection{Hurwitz numbers} 
Starting from \cite{Okounkov-2000},
expressions containing sums over $\lambda$ each term of which consists products of the Schur functions
labeled with the same partition
were used to generate
Hurwitz numbers, see for instance \cite{AMMN-2011}, \cite{Alexandrov}, \cite{AMMN-2014}, \cite{HO-2014}, 
\cite{NO-LMP}.
One can assign the 'Euler characteristic' to such sums \cite{O-2017}, by assigning $deg$ equal to 1 to 
each Schur function and getting the degree of ratios of the Schur functions. 

The present case is characterized by the fact that, firstly, Euler's characteristic can be any integer 
not exceeding 2, and secondly, an amazing coincidence of the  Euler characteristic of the base surface for 
Hurwitz numbers and the Euler characteristics of the network chord diagram (in case of orientable base surface).

By Corollary \ref{Prop2} we obtain

\begin{Theorem}\label{Theorem-2} For a given set of partitions $\mu^1=(\mu^1_1,\mu^1_2,\dots)$, $\cdots$ 
$,\mu^F=(\mu^F_1,\mu^F_2,\dots )$ the spectral correlation functions generates Hurwitz numbers as follows:
\[
\mathbb{E}_{n,N}({\bf p}_{\mu^1}(X_1)\cdots {\bf p}_{\mu^F}(X_F))
 \prod_{a=1}^F \frac{1}{z_{\mu^a}} 
  =
 \]
 \be\label{Th2}
 \delta(\mu^1,\dots,\Delta^V) N^{-nd}
\sum_{\Delta^1,\dots,\Delta^{V}} 
 H_{F-n+V}\left(\mu^1,\dots\mu^F,\Delta^1,\dots,\Delta^{V}\right)
 \prod_{i=1}^{V} {\bf p}_{\Delta^i}({\tilde C}_i)
 \ee
 where $\delta(\mu^1,\dots,\Delta^V)=1$ if $|\mu^1|=\cdots =|\mu^F|=|\Delta^1|=\cdots =|\Delta^{V}|=d$
 and vanishes otherwise.
 Here $V$ is the number of vertices of the ribbon graph, $n$ is the number of edges,
 $F$ is the number of faces.
 
In particular, if all sources matrices are equal to $\mathbb{I}_N$ we get
\be\label{Th2-I_N-lhside}
 N^{\ell(\mu^1)+\cdots +\ell(\mu^F) }
 \,\mathbb{E}_{n,N}({\bf p}_{\mu^1}(X_1)\cdots {\bf p}_{\mu^F}(X_F))
 \prod_{a=1}^F \frac{1}{z_{\mu^a}} 
  =
 \ee
 \be\label{Th2-I_N}
\sum'_{\textsc{e}} 
 H_{F-n+V}^{\textsc{e}}\left(\mu^1,\dots\mu^F; V+F\right)
 N^{\textsc{e}}
 \ee
 where the summation range is 
 $ \sum_{a=1}^F \ell(\mu^a) - nd \le \textsc{e}\le Vd+\sum_{a=1}^F \ell(\mu^a) -nd $.
\end{Theorem}

In particular,
\begin{Corollary}\label{Theorem1} 
We have
 \[
 \mathbb{E}_{n,N}({\bf p}_{\mu}(X)):=\mathbb{E}_{n,N}\left(\tr X^{\mu_1} \cdots \tr X^{\mu_\ell}\right)
 \]
 \be\label{Th1}
 ={z_\mu} N^{-nd} \sum_{\Delta^1,\dots,\Delta^{V}} 
 H_{F-n+V}\left(\mu,\Delta^1,\dots,\Delta^{V}\right)
 \prod_{i=1}^{V} {\bf p}_{\Delta^i}({\tilde C}_i)
 \ee
 where $V=n+1-2g^*=\textsc{e}^*+n-F$.
  In particular, if all ${\tilde C}_i=\mathbb{I}_N$, we get
  \[
 \mathbb{E}_{n,N}({\bf p}_{\mu}(X)):=\mathbb{E}\left(\tr X^{\mu_1} \cdots \tr X^{\mu_\ell}\right)
 \]
 \be\label{Th1'}
 ={z_\mu}N^{-nd} \sum_{g}  H^{2-2g}_{2-2g^*}\left(\mu; V+1\right)
 N^{V} 
 \ee
 where $V=n+1-2g^*$, 
 and where $H^{2-2g}_{2-2g^*}\left(\mu; V\right)$ is the Hurwitz number counting $d=|\mu|$-sheeted
 covers of Riemann surface of genus $g^*$ by Riemann surfaces of genus $g$ with at most $V+1$ critical points
(see (\ref{H-E-E^*})) for the notation $H^{\textsc{e}}_{\textsc{e}^*}$).   
\end{Corollary}

Thus, the expectation in the  r. h. s. of (\ref{Th1}) is expressed in terms of the Hurwitz numbers
which enumerate $d$-sheeted coverings of Rieman surfaces of Euler characteristic $2-2g^*$ with $n+1-g^*$ 
branch points with profiles $\mu,\Delta^1,\dots,\Delta^{n-2g^*}$
where $d=|\mu|=|\Delta^1|=\cdots=|\Delta^{n-2g^*}|$.


We get

\subsection{Non-orientable case. Hurwitz numbers for Klein surfaces.}

To get Hurwitz numbers as expectation values of spectral function we use the ``Mebius'' tau function
(\ref{tau1--hypB}):
\[
\tau_1^{\rm B}(Z):= \sum_{\lambda}s_\lambda(Z)=
\prod_{i < j}(1-z_iz_j)^{-1}\prod_{i=1}^N (1-z_i)^{-1}
\]
where $z_i,\,i=1,\dots,N$ are eigenvalues of $Z$. This trick was done in \cite{NO-2014} and \cite{O-TMP-2017}.
This tau function was pointed out in \cite{OST-I} as the simplest example of the BKP tau function.

Straightforward 
generalization of \ref{Theorem-1} reads as

\begin{Theorem}\label{Theorem-1-Klein} Under conditions of Theorem \ref{Theorem-1}  we have
 \be\label{Th-1-Klein}
\mathbb{E}_{n,N}\left( \prod_{a=1}^{F-e} \tau_{r^{(a)}}(x,X_a,N{\bf p}^{(a)}))
\prod_{a=F-e +1}^{F} \tau_{r^{(a)}}^{\rm B}(x,X_a)
\right) =
\ee
\be\label{Theor-1-Klein-Schur}
\sum_{\lambda\atop \ell(\lambda)\le N} r_\lambda(x) \left(  
s_\lambda(N{\bf p}_\infty) 
\right)^{-n} \prod_{i=1}^{F-e} s_\lambda(N{\bf p}^{(a)})
\prod_{b=1}^{V}  s_\lambda({\tilde C}_b) 
 \ee
 where $F-e>0$ and where
 \[
 r_\lambda(x):= \prod_{(i,j)\in\lambda} r(x+j-i)
 \]
 where each $\tau_{r^{(a)}}^{\rm B}$ is defined by (\ref{tau--hypB}) and where $r=\prod_{a=1}^{F} r^{(a)}$.
 The degree of the (\ref{Theor-1-Klein-Schur}) is equal to $F-e - n +V$ is equal to 
 $\textsc{e}^*-e$
 where $\textsc{e}^*$ is the Euler characteristic of the network chord diagram.
\end{Theorem}

We need $F-e>0$ to have a non-empty set of parameters ${\bf p}^{(a)}$ to provide the convergency of the 
expectation value (see Remark 1 after Theorem \ref{Theorem-1}).

One can interpret the degree $\textsc{e}^*-e$ of the (\ref{Theor-1-Klein-Schur}) as follows. The faces 
$X_1,\dots,X_{F-e}$
related to the tau functions $\tau_{r^{(a)}},\, a=1,\dots , F-e$ (let us call them punctured one) are 
treated as before. The faces $X_{F-e+1},\dots,X_F$ related to functions $\tau_r^{\rm B}$ should be 
interpreted as holes glued by Mobius sheets. Insertion of each Mobius sheet diminishes the Euler 
characteristic of the base surface by 1. This rule sounds more like mnemonic since there is yet no
direct connection of the series of the ratios of the Schur functions to the topology of surfaces.

In certain cases the expression (\ref{Theor-1-Klein-Schur}) is a tau function, see Examples 3' and 4' below, however
these cases are related to $\textsc{e}^*=1$.

Take $r=1$ below.
The analgues of Examples 3 and 4 may be chosen as

{\bf Example 4'}. Take  $F=2$, $e=1$ and $X_1=Z_1C_1\dots Z_nC_n$, $X_2=Z_n^\dag C_n^* \cdots Z_1^\dag C_1^*$. 
As one can see in this case $V=n$ (thus, $\textsc{e}^*=F-1-n+V=1$) and the words are $C_iC_i^*,\,i=1,\dots,n$.
The right hand side of (\ref{Theor-1-Klein-Schur}) is equal
to 
$$
\sum_\lambda  s_\lambda(N{\bf p}^{(1)}) \prod_{i=1}^{n} 
\frac{s_\lambda(C_iC_{i+1}^*)}{s_\lambda(N{\bf p}_\infty)},\quad C_{n+1}^*:=C_1^*
$$
In case all source matrices were $\mathbb{I}_N$, it is equal to 
$\sum_\lambda \left( (N)_\lambda \right)^{n} s_\lambda(N{\bf p}^{(1)})  $
which is the BKP tau function (\ref{tau--hypB}) with $r(x)=x^n$.

{\bf Example 5'}. Take $F=n$ and $X_a=Z_aC_aZ_{a+1}^\dag C_{a+1}^*,\,1\le a < n$ and 
$X_F=Z_nC_nZ_1^\dag C_1^*$
(a closed chain). We obtain two vertices (so, $\textsc{e}^*=2$) and two words
$\tilde{C}_1=C_1C_2\cdots C_n$, $\tilde{C}_2=C_n^*C_{n-1}^*\cdots C_1^*$. In case all source matrices
were $\mathbb{I}_N$, we get that the right hand side of (\ref{Pr2}) is equal to
$$
\sum_\lambda \left( (N)_\lambda \right)^{2} \left(s_\lambda(N{\bf p}_\infty)\right)^{1+e-n}
\prod_{a=1}^{n-e} s_\lambda(N{\bf p}^{(a)}). 
$$ 
It can be identified with the tau function (\ref{tau--hyp}), if we 
fix each set ${\bf p}^{(a)},\,a=1,\dots,n$  to be in the form (\ref{special-p}), with the exception of the selected
one that we will interpret as higher times of the BKP hierarchy.

\paragraph{Hurwitz numbers.} We get the following generation functions of Hurwitz numbers of Klein surfaces: 

\begin{Theorem}\label{Theorem-2-Klein} We have
\[
 \mathbb{E}_{n,N}\left({\bf p}_{\mu^1}(X_1)\cdots {\bf p}_{\mu^{F-e}}(X_{F-e})
 \tau^{\rm B}(X_{F-e+1})\cdots \tau^{\rm B}(X_F)\right)
  \prod_{a=1}^{F-e} \frac{1}{z_{\mu^a}}= 
 \]
 \be\label{Th2-Klein}
 \delta(\mu^1,\dots,\Delta^V)N^{-nd}
\sum_{\Delta^1,\dots,\Delta^{V}} 
 H_{\textsc{e}^*-e}\left(\mu^1,\dots\mu^{F-e},\Delta^1,\dots,\Delta^{V}\right)
 \prod_{i=1}^{V} {\bf p}_{\Delta^i}({\tilde C}_i)
 \ee
where $\delta(\mu^1,\dots,\Delta^V)=1$ if $|\mu^1|=\cdots =|\mu^{F-e}|=
|\Delta^1|=\cdots =|\Delta^{V}|$
 and vanishes otherwise.
 Here $V$ is the number of vertices of the ribbon graph (fatgraph) obtained
 from the original network, $F-e$ is the number of punctured faces.

\end{Theorem}

\subsection{Discrete ensembles, $\beta$-ensembles (not finished)}

Sums in the right hand sides of (\ref{Th-1-in-Schur-series}) and more generally of (\ref{Theor-1-Klein-Schur}) 
may be treated as discrete ensembles which
generalize known ensembles which can be related to $\textsc{e}^*$ series in the Schur functions 
\cite{KMMM} and \cite{OShiota-2004}. 

$\,$

\paragraph{$\beta$-ensemble.}
The matrix models labeled with networks may written as discrete $\beta$-ensembles if we fix parameters
${\bf p}^{(a)}$ with the help (\ref{special-p}) that means that we study expectation value of products
of powers of determinants (and one of this power should be a natural number, see Remark 1 after Theorem 
\ref{Theorem-1}). This topic will be developed 
in a more detailed version, now, let me explain the idea. One need to use relations
\be
s_\lambda(N{\bf p}(\texttt{d},a))= a^{|\lambda|}(-N\texttt{d})_\lambda
s_\lambda({\bf p}_\infty) ,
\quad s_\lambda(N{\bf p}_\infty)= N^{|\lambda|}
\frac{\prod_{i<j}^N(h_i-h_j)}{\prod_{i=1}^N h_i!}
\ee
where $h_i=\lambda_i-i+N$ are shifted parts of $\lambda$ and the notation $(-\texttt{d})_\lambda $
was defined in (\ref{Poch}). Let $N\texttt{d}_1=NL>0$ is integer.
Notice that $(-NL)_\lambda $ vanishes
for $\lambda_1>NL$. For $N\texttt{d}_i$ that are not natural numbers, we use
\[
 (-N\texttt{d}_i)_\lambda =\prod_{j=1}^{N-1} (-N\texttt{d}_i-j)^{N-j+1} 
 \prod_{j=1}^{N} \frac{\Gamma(h_j+1-N-N\texttt{d}_i)}{\Gamma(-N\texttt{d}_i)}
\]
Then, choosing any $e$ within $0 \le e \le F-1$,  we get
\[
 \mathbb{E}_{n,N}\left(  
 \det \left(1-a_1X_1\right)^{NL} 
 \prod_{i=2}^{F-1-e}
 \left(1-a_iX_i\right)^{N\texttt{d}_i}
 \prod_{i=F-e+1}^F\tau_1(X_i)
 \right)=
\]
\be\label{discrete-beta-ensemble}
=\,\frac{c_N}{N!}\,\sum_{h_1,\dots,h_N\ge 0}' \,\prod_{a<b}^N\, |h_a-h_b|^{F-n+V-e}\,
\prod_{j=1}^N \,\frac{a_1^{h_j}\prod_{i=1}^{F-e-1} a_i^{h_j}
\Gamma(h_j+1-N-N\texttt{d}_i)}{\left(\Gamma(h_j+1)\right)^{F-n+V}\Gamma(NL+N-h_j)}
\ee
where $\Sigma'$ means that all $h_i,\,i=1,\dots,N$ are different (therefore the Vandermond product
does not vanish), and
where $c_N=c_N(\{a_i,\texttt{d}_i\})=\prod_{i=1}^{F-e}
a_i^{...}
\prod_{j=1}^N 
\frac{(-N\texttt{d}_i-j)^{N-j+1}}
{\Gamma(-N\texttt{d}_i)}....$.

We intentionaly separate the case $N\texttt{d}_1=NL$  to avoid possible
divergence in the summation, with $L$ be a natural number the right hand side (\ref{discrete-beta-ensemble})
it is a finite sum with the summation range $ 0 \le h_i \le NL+N,\,i=1,\dots,N$.

One could write down the equation for the equilibrium Young diagram related to the discrete 2D Coulomb gas 
on the semiline (in case $\beta=1,2$) or, 2D 'gravitational' gas on the semiline in case $\beta <0 $.

$\,$

\paragraph{Coupled, or, Kontsevich-type ensembles.}
It may be available to fix ${\bf p}^{(2)}$ in different way as ${\bf p}^{(2)}={\bf p}^{(2)}(Y)$ where
$Y_{ij}=\delta_{i,j}\exp y_i,\,i=1,\dots, N$ (see (\ref{p(X)}) for the notation). 
The matrix $Y$ plays the role of an additional source matrix similar to the role of
external matrix in the coupled matrix model.
(One can still take any of $N\texttt{d}_i$ to be an natural number in case the sum is divergent).
Instead of (\ref{discrete-beta-ensemble}) we get
\[
 \mathbb{E}_{n,N}\left(  
 \det \left(1-Y \otimes X_1\right)^{-N} 
 \prod_{i=2}^{F-1-e}
 \left(1-a_iX_i\right)^{N\texttt{d}_i}
 \prod_{i=F-e+1}^F\tau_1(X_i)
 \right)=
\]
\be\label{discrete-kontsevich-ensemble}
\frac{\tilde{c}_N}{N!}\,\sum_{h_1,\dots,h_N\ge 0}' \,\prod_{a<b}^N\, (h_a-h_b)^{F-n+V-e-1}\,
\prod_{j=1}^N \,\frac{a_1^{h_j}\prod_{i=1}^{F-e-1} a_i^{h_j}
\Gamma(h_j+1-N-N\texttt{d}_i)}{\left(\Gamma(h_j+1)\right)^{F-n+V}
\exp (-N y_j h_j)}
\ee
where $\tilde{c}_N=c_N \prod_{i<j} (e^{Ny_i}-e^{Ny_j})$, compare to the similar replacement in \cite{KMMM}
and \cite{OShiota-2004}.

\subsection{Products of unitary matrices.\label{Product of unitary...}}
  If we replace $n$ independent complex Ginibre ensembles by $n$ independent circular
  $\beta=2$ ensembles, namely, if each $N\times N$ matrix $Z_i$ is replaced by an
  $N\times N$ unitary matrix $U_i$, and, respectively, each $Z_i^\dag$ is replaced by
  $U_i^\dag$, and the sources matrices $C_i,C_i^*$ are unitary (or, more general, matrices 
  diagonalizable by unitary transform) then we get the same Theorems \ref{Theorem-1},\ref{Theorem-1-Klein}
  where $s_\lambda(N{\bf p}_\infty)$ is replaced by $s_\lambda(\mathbb{I}_N)$, where $\mathbb{I}_N$ is
  $N\times N$ identity matrix. We also get certain versions of Theorems \ref{Theorem-2}, \ref{Theorem-2-Klein},
  however formulations of these ones needs more space (see for instance cases related to $\textsc{e}^*$ 
  in \cite{HO-2014}  and $\textsc{e}^*=1$ in \cite{NO-LMP}).
  
  For instance, the analogue of the Corollary \ref{Prop2} reads as
  
\begin{Proposition}\label{Prop2-U} 
Consider the product $X=X_1\cdots X_F$ where each matrix from the set 
$\{U_iC_i,\,U_i^\dag C_i^*,\, i=1,\dots,n \}$ enters
as a factor to the product $X$ only once.
Denote the genus of the related chord diagram $g^*$, and related words ${\tilde C}_i,\,i=1,\dots,V$,
the number of faces of the related ribbon graph (embedded graph) is equal to $F$, the number of edges
is $n$, the number of vertices is $V$, and the genus $g^*$ is defined by $2-2g^*=V-n+F $.
Then we have
 \be\label{Th-1-Schur-U}
\mathbb{E}_{n,N}\left( e^{\tr \texttt{V}(X_1,{\bf p}^{1})}\cdots e^{\tr \texttt{V}(X_F,{\bf p}^{F})} \right) =
\sum_{\lambda}\left(  
s_\lambda({ \mathbb{I}_N}) 
\right)^{-n} \prod_{i=1}^F s_\lambda(N{\bf p}^{(i)})
\prod_{i=1}^{V}  s_\lambda({\tilde C}_i)
 \ee
 In particular, if all source matrices are equal to $\mathbb{I}_N$ and ${\bf p}^{(i)}
={\bf p}^{(i)}(\texttt{d}_i,a_i)$ (namely, $p_m^{(i)}=-\texttt{d}_ia_i^m$, $m>0$) we get
\be
\mathbb{E}_{n,N}\left( \prod_{i=1}^F\det\left(1-a_iX_i \right)^{N\texttt{d}_i} \right) =
\sum_{\lambda}\left(  s_\lambda({ \mathbb{I}_N}) 
\right)^{V-n+F} \prod_{i=1}^F \frac{(N\texttt{d}_i)_\lambda}{(N)_\lambda}
\ee

\end{Proposition}

\subsection{The sketch of proofs.}

First,
we know how to evaluate the integrals with the Schur function via Lemma
used in \cite{O-2004-New} and \cite{NO-2014,NO-LMP}
(for instance see \cite{Mac} for the derivation). 
\bl \label{useful-relations}
Let $A$ and $B$ be normal  matrices (i.e. matrices diagonalizable by unitary transformations). Then
\begin{equation}\label{sAUBU^+1}
\int_{\mathbb{U}(N)}s_\lambda(AUBU^{-1})d_*U=
\frac{s_\lambda(A)s_\lambda(B)}{s_\lambda(\mathbb{I}_N)} \ ,
\end{equation}
For $A,B\in GL(N)$ we have
\begin{equation}\label{sAUU^+B'}
\int_{\mathbb{U}(n)}s_\mu(AU)s_\lambda(U^{-1}B)d_*U=
\frac{s_\lambda(AB)}{s_\lambda(\mathbb{I}_N)}\delta_{\mu,\lambda}\,.
\end{equation}
Below ${\bf p}_{\infty}=(1,0,0,\dots)$. 
\begin{equation}\label{sAZBZ^+'}
\int_{\mathbb{C}^{N^2}} s_\lambda(AZBZ^+)\text{e}^{-N\operatorname{tr}
	ZZ^+}\prod_{i,j=1}^N d^2Z_{ij}=
\frac{s_\lambda(A)s_\lambda(B)}{s_\lambda(N{\bf p}_{\infty})}
\end{equation}
and
\begin{equation}\label{sAZZ^+B'}
\int_{\mathbb{C}^{N^2}} s_\mu(AZ)s_\lambda(Z^+B) \text{e}^{-N\operatorname{tr}
	ZZ^+}\prod_{i,j=1}^N d^2Z_{ij}= \frac{s_\lambda(AB)}{s_\lambda(N{\bf p}_{\infty})}\delta_{\mu,\lambda}\,.
\end{equation}
\el

These relations are used for step-by-step integration (Gaussian in the case of complex matrices).

As we can see, these relations perform the procedure of cutting and joining loops in a network of chord diagrams, 
and also create edges of ribbon graph (each edge is a coupled pair of conjugate random matrices). 
Namely, the equation (\ref{sAZBZ^+'}) performs the splitting of the loop $AZBZ^\dag$ into two loops,
$A$ and $B$,
for complex Ginibre ensembles (the resulting equation (\ref{sAUBU^+1}) splits the loop $AUBU^\dag $ 
for circular ensembles), and equation
(\ref{sAZZ^+B'}) performs the union of two loops $A$ and $B$ for complex Ginibre ensembles  
(and the equation (\ref{sAUU^+B'}) does the same for circular
ensembles). Every time we apply some of the relations (\ref{sAUU^+B'})-(\ref{sAZZ^+B'}), 
we get the factor (the "propagator" of the edge of the ribbon graph), which is 
$ \frac{1}{s_\lambda(N{\bf p}_\infty)} $
in the case of complex Ginibre ensemble  and $ \frac{1}{s_\lambda(\mathbb{I}_N)} $ in the circular case.

In this way we prove Theorem \ref{Theorem-1} and Theorem \ref{Theorem-1-Klein} and their
analogues for the circular ensembles.

Then, Theorems \ref{Theorem-2} and \ref{Theorem-2-Klein} follows, respectively, from Theorems \ref{Theorem-1}
and Theorem \ref{Theorem-1-Klein}, and by the usage of
the Mednykh formula (\ref{MednykhHurwitz}) and the characteristic map relation
\cite{Mac}:
\be\label{Schur-char-map}
s_\lambda(N\bpow)=
\frac{\operatorname{dim}\lambda}{d!}\,\sum_{\Delta\atop |\Delta|=|\lambda| } 
\varphi_\lambda(\Delta)\bpow_{\Delta}N^{\ell(\Delta)}
\ee
where $\ell(\Delta)$ is the length of the partition $\Delta$, where
$\bpow_\Delta=p_{\Delta_1}\cdots p_{\Delta_{\ell}}$ and where the summation ranges over
all partitions $\Delta=(\Delta_1,\dots,\Delta_\ell)$ whose weight
coinsides with the weight of $\lambda$: $|\lambda|=|\Delta|=d$. Here 
\be
\operatorname{dim}\lambda =d!s_\lambda(\bpow_\infty),\qquad \bpow_\infty = (1,0,0,\dots)
\ee
is the dimension of the irreducable representation of the symmetric group $S_d$. We imply that 
$\varphi_\lambda(\Delta)=0$ if $|\Delta|\neq |\lambda|$.

\section*{Acknowledgements}

This work was done during my visits to Belowezie, Bialystok and to Anger university. I am grateful to director 
of the Institute of Mathematics in Bialystok Anatol Odzijewicz and to Prof. Vladimir Roubtsov
in Anger university for their kind hospitality.
The work has been funded by RFBR grant 18-01-00273a and the RAS Program ``Fundamental problems of nonlinear mechanics'' and 
by the Russian Academic Excellence Project~\mbox{'5-100'}.
I thank Borot, A.Mudrov, S.Lando and M.Kazarian for their remarks which allow me to compare my results
with combinatorial problems already appeared in the well-known model of Hermitian matrices,
and also to E. Strakhov who drew my attention to the works on quantum chaos devoted to the products 
of random matrices and for fruitful discussions. I thankful to S.Natanzon, A.Mironov and J.Harnad
for numerous discussions on Hurwitz numbers.

\appendix

\section{Counting of branched covers \label{Counting-of-branched-covers}}

In this section the Euler characteristic of the base surface is denoted $\textsc{e}$.

Let us consider a connected compact surface without boundary $\Omega$ and a branched covering $f:\Sigma\rightarrow\Omega$
by a connected or non-connected surface $\Sigma$. We will consider a covering $f$ of the degree $d$. It means that the
preimage $f^{-1}(z)$ consists of $d$ points $z\in\Omega$ except some finite number of points. This points are called
\textit{critical values of $f$}.

Consider the preimage $f^{-1}(z)=\{p_1,\dots,p_{\ell}\}$ of $z\in\Omega$. Denote by $\delta_i$ the degree of $f$ at 
$p_i$. It means that in the neighborhood of $p_i$ the function $f$ is homeomorphic to $x\mapsto x^{\delta_i}$. The set 
$\Delta=(\delta_1\dots,\delta_{\ell})$
is the partition of $d$, that is called \textit{topological type of $z$}.

For a partition $\Delta$ of a number $d=|\Delta|$ denote by $\ell(\Delta)$ the number of the non-vanishing parts 
($|\Delta|$ and $\ell(\Delta)$ are called the weight and the length of $\Delta$, respectively). We denote a partition 
and its Young diagram by the same letter. Denote by $(\delta_1,\dots,\delta_{\ell})$ the Young diagram with rows of l
ength $\delta_1,\dots,\delta_{\ell}$ and corresponding partition of $d=\sum \delta_i$.

Fix now points $z_1,\dots,z_k$ and partitions $\Delta^{(1)},\dots,\Delta^{(k)}$ of $d$. Denote by
\[\widetilde{C}_{\Omega (z_1\dots,z_{k})} (d;\Delta^{(1)},\dots,\Delta^{(k)})\]
the set of all branched covering $f:\Sigma\rightarrow\Omega$ with critical points $z_1,\dots,z_{k}$ of 
topological types  $\Delta^{(1)},\dots,\Delta^{(k)}$.

Coverings $f_1:\Sigma_1\rightarrow\Omega$ and $f_2:\Sigma_2\rightarrow\Omega$ are called isomorphic if there exists an
homeomorphism $\varphi:\Sigma_1\rightarrow\Sigma_2$ such that $f_1=f_2\varphi$. Denote by $\texttt{Aut}(f)$  the order 
of the group of automorphisms of the covering $f$. Isomorphic coverings have isomorphic groups of automorphisms of 
degree $|\texttt{Aut}(f)|$.

Consider now the set $C_{\Omega (z_1\dots,z_{k})} (d;\Delta^{(1)},\dots,\Delta^{(k)})$ of isomorphic 
classes in $\widetilde{C}_{\Omega (z_1\dots,z_{k})} (d;\Delta^{(1)},\dots,\Delta^{(k)})$. This is a 
finite set.
The sum
\be\label{Hurwitz-number-geom-definition}
H_{\textsc{e}}(\Delta^{(1)},\dots,\Delta^{(k)})=
\sum\limits_{f\in C_{\Omega (z_1\dots,z_{k})}(d;\Delta^{(1)},\dots,
	\Delta^{(k)})}\frac{1} {|\texttt{Aut}(f)|}\quad,
\ee
don't depend on the location of the points $z_1\dots,z_{k}$ and is called \textit{Hurwitz number}.
Here $k$ denotes the number of the branch points, and $\textsc{e}$ is the Euler characteristic of the base surface.

In case it will not produce a confusion we admit 'trivial' profiles $(1^d)$ 
among $\Delta^1,\dots,\Delta^{k}$ in
(\ref{Hurwitz-number-geom-definition})
keeping the notation $H_{\textsc{e}}(\Delta^{(1)},\dots,\Delta^{(k)})$ 
though the number of critical points now is less than $k$.

In case we count only connected covers $\Sigma$ we get the \textit{connected} Hurwitz numbers 
$H^{\rm con}_{\textsc{e}^*}(\Delta^{(1)},\dots,\Delta^{(k)})$.

\vspace{1ex}

The Hurwitz numbers arise in different fields of mathematics: from algebraic geometry to integrable systems. 
A special interest in this topic arose after the papers \cite{Dijkgraaf} and \cite{ELSV} (see \cite{KazarianLando}
and \cite{ZL} for a review).
They are 
well studied for orientable $\Omega$. In this case the Hurwitz number coincides with the weighted number of holomorphic
branched coverings of a Riemann surface $\Omega$ by other Riemann surfaces, having critical points 
$z_1,\dots,z_k\in\Omega$ of the topological types $\Delta^{(1)},\dots,\Delta^{(k)}$ respectively. 
The well known isomorphism between Riemann surfaces and complex algebraic curves gives the interpretation of the 
Hurwitz numbers as the numbers of morphisms of complex algebraic curves.

Similarly, the Hurwitz number for a non-orientable surface $\Omega$ coincides with the weighted number of the 
dianalytic branched coverings of the Klein surface without boundary by another Klein surface and coincides with the 
weighted number of morphisms of real algebraic curves without real points \cite{AG,N90,N2004}. An extension of the 
theory to all Klein surfaces and all real algebraic curves leads to Hurwitz numbers for surfaces
with boundaries may be found in \cite{AN,N}.

Riemann-Hurwitz formula related the Euler characteristic of the base surface $\textsc{e}^*$ and the Euler 
characteristic of the $d$-sheeted cover $\textsc{e}$ as follows:
\be\label{RH}
\textsc{e}= d\textsc{e}^* +\sum_{i=1}^{k}\left(\ell(\Delta^{(i)})-d\right)=0
\ee
where the sum ranges over all branch points $z_i\,,i=1,2,\dots$ with ramification profiles given by partitions 
$\Delta^i\,,i=1,2,\dots$ respectively, and $\ell(\Delta^{(i)})$ denotes the length of the partition $\Delta^{(i)}$ 
which is equal to the number of the preimages $f^{-1}(z_i)$ of the point $z_i$.
\begin{examp}
Let $f:\Sigma\rightarrow\mathbb{CP}^1$ be a covering without critical points.
Then, each $d$-sheeted cover is the disjoint union of $d$ Riemann spheres: 
$\mathbb{CP}^1 \coprod \cdots \coprod \mathbb{CP}^1$, then
$|{\rm Aut} f|  =d!$ and $H_{2}((1^d))=\frac{1}{d!}$. The same answer one gets from Mednykh formula 
(\ref{MednykhHurwitz}).
\end{examp}
\begin{examp}
Let $f:\Sigma\rightarrow\mathbb{CP}^1$ be a $d$-sheeted covering with two critical points with the profiles 
$\Delta^{(1)}=\Delta^{(2)}=(d)$.
(One may think of $f=x^d$). Then $H_{2}((d),(d))=\frac 1d$. Let us note that $\Sigma$ is connected in this case 
(therefore $H_{2}((d),(d))=H^{\rm con}_{2}((d),(d)) $)
and its Euler
characteristic $\textsc{e}=2$.
\end{examp}
\begin{examp}
The generating
function for the Hurwitz numbers $H_{2}((d),(d))$ from the previous Example may be writen as
$$
F(h^{-1}\bpow^{(1)},h^{-1}\bpow^{(2)}):=\, h^{-2}\sum_{d>0}\, H^{\rm con}_{2}((d),(d)) p_d^{(1)}p_d^{(2)}=
h^{-2}\sum_{d>0} \frac 1d p_d^{(1)}p_d^{(2)}
$$ 
Here $\bpow^{(i)}=(p_1^{(i)},p_2^{(i)},\dots),\,i=1,2$ are two sets of formal parameters. The powers of the auxilary
parameter $\frac 1h$ count the Euler characteristic of the cover $\textsc{e}$ which is 2 in our example.
Then thanks to the known general statement about the link between generating functions of ``connected'' and ``disconnected''
Hurwitz numbers (see for instance \cite{ZL}) one can write down the generating function for the Hurwitz numbers for covers with
two critical points,
$H_{2}(\Delta^{(1)},\Delta^{(2)})$, as follows:
\begin{multline}\label{E=2,F=2Hurwitz}
\tau(h^{-1}\bpow^{(1)},h^{-1}\bpow^{(2)})=\text{e}^{F(h^{-1}\bpow^{(1)},h^{-1}\bpow^{(2)}) }\\ =
\text{e}^{h^{-2}\sum_{d>0} \frac 1d p_d^{(1)}p_d^{(2)}}\,
= \,
\sum_{d\ge 0} \sum_{\Delta^{(1)},\Delta^{(2)}}
H_{2}(\Delta^{(1)},\Delta^{(2)}) \,h^{-\textsc{e}'} \bpow^{(1)}_{\Delta^{(1)}}\bpow^{(2)}_{\Delta^{(2)}}
\end{multline}
where $\bpow^{(i)}_{\Delta^{(i)}}:=
p^{(i)}_{\delta^{(i)}_1}p^{(i)}_{\delta^{(i)}_2}p^{(i)}_{\delta^{(i)}_3}\cdots$, $i=1,2$
and where $\textsc{e}=\ell(\Delta^{(1)}) + \ell(\Delta^{(2)})$ in agreement with (\ref{RH}) 
where we put $k=2$.
From (\ref{E=2,F=2Hurwitz}) it follows that the profiles of both critical points
coincide, otherwise the Hurwitz number vanishes. Let us denote this profile $\Delta$, 
and $|\Delta|=d$ and from the last equality we get
$$
H_{2}(\Delta,\Delta) = \frac {1}{z_{\Delta}}
$$
Here
\be
z_\Delta\,=\,\prod_{i=1}^\infty \,i^{m_i}\,m_i!
\ee
where $m_i$ denotes the number of parts equal to $i$ of the partition $\Delta$ 
(then the partition $\Delta$ is often
denoted by $(1^{m_1}2^{m_2}\cdots)$). 
\end{examp}
\begin{examp}\label{RP^2=no-crit-point}
Let $f:\Sigma\rightarrow\mathbb{RP}^2$ be a covering without critical points.
Then, if $\Sigma$ is connected, then $\Sigma=\mathbb{RP}^2$,
$\deg f=1$\quad or $\Sigma=S^2$, $\deg f=2$. Next, if $d=3$, then
$\Sigma=\mathbb{RP}^2\coprod\mathbb{RP}^2\coprod\mathbb{RP}^2$ or $\Sigma=\mathbb{RP}^2\coprod S^2$.
Thus, $H_{1}((1^3))=\frac{1}{3!}+\frac{1}{2!}=\frac{2}{3}$. The same answer one obtains from the combinatorial 
definition. Indeed, the equation $R^2=1$ has 4 solutions in $S_3$.
\end{examp}
\begin{examp}
Let $f:\Sigma\rightarrow\mathbb{RP}^2$ be a covering with a single critical point with profile $\Delta$, and $\Sigma$ 
is connected.
Note that due to (\ref{RH}) the Euler
characteristic of $\Sigma$ is $\textsc{e}'=\ell(\Delta)$. 
(One may think of $f=z^d$ defined in the unit disc where we identify $z$ and $-z$ if $|z|=1$).
In case we cover the Riemann sphere by the Riemann sphere $z\to z^m$ we get
two critical points with the same profiles. However we cover $\mathbb{RP}^2$ by the Riemann sphere, then we have the 
composition of the
mapping $z\to z^{m}$ on the
Riemann sphere and the factorization by antipodal involution $z\to - \frac{1}{\bar z}$. Thus we have the ramification 
profile $(m,m)$
at the single critical point $0$ of $\mathbb{RP}^2$.
The automorphism group is the dihedral group of the order $2m$ which consists of rotations on $\frac{2\pi }{m}$ and 
antipodal involution
$z\to -\frac{1}{\bar z}$.
Thus we get that 
$$
H^{\rm con}_{1}\left(2m;(m,m)\right)=\frac{1}{2m}
$$
From (\ref{RH}) we see that $1=\ell(\Delta)$ in this case.
Now let us cover $\mathbb{RP}^2$ by $\mathbb{RP}^2$ via $z\to z^d$. From (\ref{RH}) we see that $\ell(\Delta)=1$.
For even $d$ we have the critical point
$0$, in addition each point of the unit
circle $|z|=1$ is critical (a folding), while from the beginning we restrict our consideration only on isolated 
critical points. For odd $d=2m-1$ there is
the single critical point $0$, the automorphism group consists of rotations on the angle $\frac{2\pi}{2m-1}$. 
Thus, in this case
$$
H^{\rm con}_{1}\left(2m-1;(2m-1)\right)=\frac{1}{2m-1}
$$
\end{examp}
\begin{examp}
The generating series of the connected Hurwitz numbers with a single critical point from the previous Example  is
\begin{multline*}
F(h^{-1}\bpow)=
\frac{1}{h^2}\sum_{m>0} p_m^2 H^{\rm con}_{1}\left(2m;(m,m)\right)\\ +
\frac{1}{h} \sum_{m>0} p_{2m-1} H^{\rm con}_{1}\left(2m-1;(2m-1)\right)
\end{multline*}
where $H^{{\rm con}}_{1}$ describes $d$-sheeted covering either by the Riemann
sphere ($d=2m$) or by the projective plane ($d=2m-1$). 
We get the generating function for  Hurwitz numbers with a single critical point
\begin{multline}\label{single-branch-point'}
\tau(h^{-1}\bpow)=\text{e}^{F(h^{-1}\bpow ) }\\ =
\text{e}^{\frac {1}{h^2}\sum_{m>0} \frac {1}{2m}p_m^2  +\frac 1h\sum_{m {\rm odd}} \frac 1m p_m }=
\sum_{d>0} 
\sum_{\Delta\atop |\Delta|=d} h^{-\ell(\Delta)} \bpow_\Delta
H_{1}(d;\Delta)
\end{multline}
 Then $H_{1}(d;\Delta)$ is the Hurwitz number 
describing $d$-sheeted covering of $\mathbb{RP}^2$ with a single
branch point of type $\Delta=(d_1,\dots,d_l),\,|\Delta|=d$ by a (not necessarily connected) Klein surface of
Euler characteristic $\textsc{e}'=\ell(\Delta)$. For instance, for $d=3$, $\textsc{e}'=1$ we get
$H_{1}(\Delta)=\frac 13\delta_{\Delta,(3)}$.
For unbranched coverings (that is for $\Delta=(1^d)$) we get the generating formula 
$e^{\frac{c^2}{2}+c}=\sum_{d\ge 0} c^d H_{1}\left(d;(1^d)\right)$.
\end{examp}

One can also get the answers considered in the examples by the usage of the Mednykh formula (\ref{MednykhHurwitz}).

\paragraph{Corollaries of the Mednykh-Pozdnyakova Character Formula \cite{NO-LMP}.}

It follows from the paper \cite{Dijkgraaf} by Dijkgraaf that the Hurwitz numbers for closed orientable 
surfaces form a 2D topological field theory.
An extension of this result to the case of Klein surfaces (thus to orientable and non-orientable 
surfaces) was found in Theorem~5.2 of \cite{AN},
(see also Corollary 3.2 in \cite{AN2008}). On the other hand, the 
Mednykh-Pozdnyakova formula describes
the Hurwitz numbers in terms of characters of the symmetric groups. One can
 interpret the axioms of
the Klein topological field theory \cite{AN} for Hurwitz numbers in terms of characters of symmetric groups.

\bl
\label{Hurwitz-down-Lemma}
\begin{eqnarray}\label{Hurwitz=Hurwirz-Hurwitz}
H_{\textsc{e}+\textsc{e}_1}(\Delta^{(1)},\dots,\Delta^{(k+k_1)})&&\\\noalign{\smallskip}
  &&\hspace{-3cm} = \sum_{\Delta}\frac{d!}{|C_\Delta|}
H_{\textsc{e}+1}(\Delta^{(1)},\dots,\Delta^{(k)},\Delta)
H_{\textsc{e}_1 +1}(\Delta,\Delta^{(k+1)},\dots,\Delta^{(k_1)})\,.\nonumber
\end{eqnarray}
In particular,
\be\label{Hurwitz-down}
H_{\textsc{e}-1}(\Delta^{(1)},\dots,\Delta^{(k)})=
\sum_{\Delta}\,
H_{\textsc{e}}(\Delta^{(1)},\dots,\Delta^{(k)},\Delta) \chi(\Delta)\,,
\ee
where $\chi(\Delta)={d!H_{1}(\Delta)}/{|C_\Delta|}$ are rational numbers explicitly 
defined in the following way by a partition $\Delta$: 
\be\label{delta(Delta)}
\chi(\Delta)=\sum_{\lambda \atop |\lambda|=|\Delta|} \chi_\lambda(\Delta)=\left[
\prod_{i>0,\,{\rm even}}e^{\frac {i}{2}\frac{\partial^2}{\partial p_i^2}}\cdot p_i^{m_i}
\prod_{i>0,\,{\rm odd}}e^{\frac {i}{2}\frac{\partial^2}{\partial p_i^2}+
\frac{\partial}{\partial p_i}}\cdot p_i^{m_i}
\right]_{\bpow =0}\,,
\ee
and $\chi_\lambda(\Delta)$ is the character of the representation $\lambda$ of the symmetric group
$S_d$, $d=|\lambda|$, evaluated on the cycle class $\Delta=(1^{m_1}2^{m_2}\cdots)$.

\el
\noindent As a corollary we get that the Hurwitz numbers of the projective plane may be obtained from
the Hurwitz numbers of the Riemann sphere, while the Hurwitz numbers of the torus and  the Klein bottle 
may be obtained from the Hurwitz numbers of the projective plane.

\paragraph{On combinatorial approach.} The study of the homomorphisms between the fundemental group of the base Riemann sufrace 
of genus $g^*$ (the Euler characterisic is resectively $\textsc{e}=2-2g^*$)
with $k$ marked points and the symmetric group in the context of the counting of the non-equivalent 
$d$-sheeted covering with given profiles 
$\Delta^{i},\,i=1,\dots,k$ results to the equation (\ref{Hom-pi-S_d-Riemann})
(for instance, for the details, see Appendix A written by Zagier for the 
Russian edition of \cite{ZL} or works \cite{M1,GARETH.A.JONES})

For instance, Example 3 considered above counts non-equivalent solutions to the equation $A_1A_2=1$ with given 
cycle classes $\textsc{C}_{\Delta^1}$ and $C_{\Delta^2}$. Solutions of this equation consist of all elements of class 
$\textsc{C}_{\Delta^1}$ and inverse elements, so $\Delta^2=\Delta^1=:\Delta$. The number of elements of any class 
$\textsc{C}_\Delta$ (the cardinality of $|\textsc{C}_\Delta|$) divided by $|\Delta|!$ is $1 \over z_\Delta$ as we 
got in the Example 3.

For Klein surfaces (see \cite{M2},\cite{GARETH.A.JONES}) instead of (\ref{Hom-pi-S_d-Riemann}) we get 
(\ref{Hom-pi-S_d-Klein}).

In (\ref{Hom-pi-S_d-Klein}),
$\textsl{g}^*$ is the so-called genus of non-orientable surface which is related to its Eular chatacteristic
$\textsc{e}^*$ as 
$\textsc{e}=2-\textsl{g}^*$. For the projective plane ($\textsc{e}^*=1$) we have $\textsl{g}^*=1$, 
for the Klein bottle ($\textsc{e}^*=1$) $\textsl{g}^*=2$.

Consider unbranched coverings ($k=0$) of the torus (equation (\ref{Hom-pi-S_d-Riemann}) where $g=1$ ), 
of the projective plane and the Klein bottle (equation (\ref{Hom-pi-S_d-Klein}) where respectively 
$\textsl{g}^*=0$ and $\textsl{g}^*=1$). For the real projective plane we have $\textsl{g}^*=1$ in 
(\ref{Hom-pi-S_d-Klein}) only one $R_0=ab$. If we treat the projective plane as the unit disk
with identfied opposit points of the boarder $|z|=1$, then $R$ is related to the path from $z$ to $-z$.
For the Klein bottle ($\textsl{g}=2$ in (\ref{Hom-pi-S_d-Klein})) there are $R_0=ba^{-1}$ and $R_1=a$.

\section{Partitions and Schur functions \label{Partitions-and-Schur-functions}}

Let us recall that the characters of the unitary group $\mathbb{U}(N)$ are labeled by partitions
and coincide with the so-called Schur functions \cite{Mac}. 
A partition 
$\lambda=(\lambda_1,\dots,\lambda_n)$ is a set of nonnegative integers $\lambda_i$ which are called
parts of $\lambda$ and which are ordered as $\lambda_i \ge \lambda_{i+1}$. 
The number of non-vanishing parts of $\lambda$ is called the length of the partition $\lambda$, and will be denoted by
$\ell(\lambda)$. The number $|\lambda|=\sum_i \lambda_i$ is called the weight of $\lambda$. The set of all
partitions will be denoted by $\mathbb{P}$.

The Schur function labelled by $\lambda$ may be defined as  the following function in variables
$x=(x_1,\dots,x_N)$ :
\be\label{Schur-x}
s_\lambda(x)=\frac{\det \left[x_j^{\lambda_i-i+N}\right]_{i,j}}{\det \left[x_j^{-i+N}\right]_{i,j}}
\ee
in case $\ell(\lambda)\le N$ and vanishes otherwise. One can see that $s_\lambda(x)$ is a symmetric homogeneous 
polynomial of degree $|\lambda|$ in the variables $x_1,\dots,x_N$, and $\deg x_i=1,\,i=1,\dots,N$.

\begin{Remark}\label{notation} In case the set $x$ is the set of eigenvalues of a matrix $X$, we also write $s_\lambda(X)$ instead
of $s_\lambda(x)$.
\end{Remark}

There is a different definition of the Schur function as quasi-homogeneous non-symmetric polynomial of degree $|\lambda|$ in 
other variables, the so-called power sums,
$\bpow =(p_1,p_2,\dots)$, where $\deg p_m = m$.

For this purpose let us introduce 
$$
s_{\{h\}}(\mathbf p)=\det[s_{(h_i+j-N)}(\mathbf p)]_{i,j},
$$
where $\{h\}$ is any set of $N$ integers, and where
the Schur functions $s_{(i)}$ are defined by $\text{e}^{\sum_{m>0}\frac 1m p_m z^m}=\sum_{m\ge 0} s_{(i)}(\bpow) z^i$.
If we put $h_i=\lambda_i-i+N$, where $N$
is not less than the length of the partition $\lambda$, then
\begin{equation}\label{Schur-t}
s_\lambda(\mathbf p)= s_{\{h\}}(\mathbf p).
\end{equation}

The Schur functions defined by (\ref{Schur-x}) and by (\ref{Schur-t}) are equal,  $s_\lambda(\bpow)=s_\lambda(x)$, 
provided the variables $\bpow$ and $x$ are related by the power sums relation
\be
\label{t_m}
p_m=  \sum_{i=1}^N x_i^m
\ee

In case the argument of $s_\lambda$ is written as a non-capital fat letter  the definition (\ref{Schur-t}),
and we imply the definition (\ref{Schur-x}) in case the argument is not fat and non-capital letter, and
in case the argument is capital letter which denotes a matrix, then it implies the definition (\ref{Schur-x}) 
with $x=(x_1,\dots,x_N)$ being the eigenvalues.

It may be easily checked that
\be\label{p-to-p-in-Schur}
s_\lambda(\bpow)=(-1)^{|\lambda|}s_{\lambda^{\rm tr}}(-\bpow)
\ee
where $\lambda^{\rm tr}$ is the partition conjugated to $\lambda$ (in \cite{Mac} it is denoted by $\lambda^*$). The Young diagram
of the conjugated partition is obtained by the transposition of the Young diagram of $\lambda$ with respect to its main diagonal. 
One gets $\lambda_1=\ell(\lambda^{\rm tr})$. And then it follows that for $L\times L$ matrix $X$ the Schur
function $s_\lambda(-{\bf p}(X))$ vanishes if $\lambda_1>L$.

\section{More about tau functions}

The product over all nodes of the Young 
diagram $\lambda$ is called {\em content product} and plays a certain role in the 
representation theory of symmetric groups (in the context of Hurwitz numbers see, for instance 
\cite{Goulden-Jackson-2008}, \cite{Harnad-2014}).
These $\tau_r$  parametrized by the choice of the function $r$ 
form the family of the Toda lattice (TL) tau functions where the sets
${\bf p}(X)$ and ${\bf p}$ play the role of the so-called higher times and the (discrete)
variable $x$ plays the role of the site number in the lattice.
The content product can be viewed as the generalized Pochhammer symbol related to Young diagrams. That's why 
such family of tau functions \cite{OS-2000} were called hypergeometric ones. 
Let us also note that TL hypergeometric functions generate Hurwitz numbers themselves, in this case
the base surface is Riemann sphere ($\textsc{e}^*=2$), see 
\cite{Goulden-Jackson-2008},\cite{AMMN-2014},\cite{HO-2014},\cite{NO-LMP}.
In addition, there exist numerous representations of TL hypergeometric tau functions
in form of matrix integrals, see, for instance \cite{O-2004-New}.

This $\tau_r^{\rm B}$ is called hypergeometric tau function \cite{OST-I} of the ``large'' BKP hierarchy, 
(BKP hierarchy introduced in \cite{KvdLbispec}).  
Similar to the Toda lattice case, $\tau_r^{\rm B}$ generates Hurwitz numbers, however in this case
$\textsc{e}^*=1$, see \cite{NO-2014},\cite{NO-LMP}.

\section{Matrix integrals as generating functions of Hurwitz numbers from \cite{NO-2014,NO-LMP}
	\label{Matrix-integrals}}

	Hurwitz numbers can be generated by series in the Schur functions. In turn, series in the Schur functions
	can be generated as perturbation series of various matrix models.
	Let us note that the very first papers devoted to the perturbation series of
	certain matrix models in terms of the Schur functions was \cite{Kazakov}.

In case the base surface is $\mathbb{CP}^1$ the set of examples of matrix integrals generating Hurwitz numbers were studied in
works \cite{Chekhov-2014,MelloKochRamgoolam,AMMN-2014,ChekhovAmbjorn,KZ,ZL,Zog}.
One can show that the perturbation series in coupling constants of these integrals (Feynman graphs) may be related to TL
(KP and two-component KP) hypergeometric tau functions. 
It actually means that these series generate Hurwitz numbers with at most two arbitrary profiles
(An arbitray profile corresponds to a certain term in the perturbation series in the coupling constants which are higher times.
The TL and 2-KP hierarchies there are two independent sets of higher times
which yeilds two critical points for Hurwitz numbers).

Here, very briefly, we will write down few generating series for the $\mathbb{RP}^2$ Hurwitz numbers.
These series may be not tau functions themselves but may be presented as integrals of tau functions of matrix argument.
(The matrix argument, which we denote by a capital letter, say $X$, means that the power sum variables $\bpow$ are specified
as $p_i=\tr X^i,\,i>0$. Then instead of
$s_\lambda(\bpow)$, $\tau(\bpow)$ we write $s_\lambda(X)$ and $\tau(X)$). If a matrix integral in examples below is a BKP tau
function then it generates Hurwitz numbers with a single arbitrary profile and all other are subjects of restrictions
identical to those in $\mathbb{CP}^1$ case mentioned above.
In all examples $\texttt{V}(x,\bpow):=\sum_{m>0} \frac 1m x^m p_m$. We also recall the notation  $\bpow_\infty=(1,0,0,\dots)$.
We also recall that numbers
$H_{\textsc{e}}(d;\dots)$ are Hurwitz numbers only in case $d\le N$, $N$ is the size of matrices.

For more details of the $\mathbb{RP}^2$ case see \cite{NO-2014}. New development in \cite{NO-2014} with respect to
the consideration in \cite{O-2004-New} is the usage of products of matrices. 
Here we shall consider a few examples. 
All examples include the simplest BKP tau function, of matrix argument $X$ written down in (\ref{tau1--hypB})
as the part of the integration measure. Other integrands are the simplest KP tau functions
$\tau_1^{\rm 2KP}(X,\bpow):=\text{e}^{\tr \texttt{V}(X,\bpow)}$ where  the parameters
$\bpow$ may be called coupling constants. The perturbation series in coupling constants are expressed
as sums of products of the Schur functions over partitions and are similar to the series we considered in
the previous sections.
\begin{exampB}
The projective analog of Okounkov's generating series for double Hurwitz series as a model of normal matrices.
From the equality
\[
\left({2\pi}{\zeta_1^{-1}} \right)^{\frac 12} \text{e}^{\frac{(n\zeta_0)^2}{2\zeta_1}} \text{e}^{\zeta_0 nc+ \frac12 \zeta_1 c^2}\,
=\,
\int_{\mathbb{R}} \text{e}^{x_i n\zeta_0 +(cx_i- \frac12 x^2_i)\zeta_1} dx_i ,
\]
in a similar way as was done in \cite{OShiota-2004} using $\varphi_\lambda(\Gamma)=\sum_{(i.j)\in\lambda}(j-i)$,
one can derive
{\small\[
\text{e}^{n|\lambda|\zeta_0}\text{e}^{\zeta_1 \varphi_\lambda(\Gamma)}\delta_{\lambda,\mu}\,=\,\textsc{k} \,
\int  s_\lambda(M) s_\mu(M^\dag) \det \left(MM^\dag\right)^{n\zeta_0}
\text{e}^{-\frac12 \zeta_1\tr \left( \log \left( MM^\dag\right)\right)^2} dM
\]}
where $\textsc{k}$ is unimportant multiplier, where $M$ is a normal matrix with eigenvalues $z_1,\dots,z_N$ and $\log |z_i|=x_i$,
and where (see \cite{MWZ})
\[dM=\,d_*U\,\prod_{i<j}|z_i-z_j|^2\prod_{i=1}^N d^2 z_i.\] Then the $\mathbb{RP}^2$ analogue of Okounkov's generating series
may be presented as the following integral
(\cite{Okounkov-2000}) may be written
\begin{multline}
\label{Okounkov-tau-normal-matrices-BKP}
\sum_{\lambda\atop \ell(\lambda)\le N}\text{e}^{n|\lambda|\zeta_0 +
	\zeta_1 \varphi_\lambda(\Gamma)}
s_\lambda(\bpow)\\
=\textsc{k}
\int  \text{e}^{\tr\texttt{V}(M,\bpow)}
\text{e}^{\zeta_0 n\tr \log \left(MM^\dag\right)-\frac12 \zeta_1\left( \tr\log \left( MM^\dag\right)\right)^2}
\tau^{\rm B}_1(M^\dag) dM
\end{multline}
Recall that in the work \cite{Okounkov-2000} there were studied Hurwitz numbers with an arbitrary number of simple branch points
and two arbitrary profiles. In our analog, describing the coverings of the projective plane, an arbitrary profile
only one, because, unlike the Toda lattice, the hierarchy of BKP has only one set of (continuous) higher times.

A similar representation of the Okounkov $\mathbb{CP}^1$   was earlier presented in
\cite{AlexandrovZabrodin-Okounkov}.

Below we use the following notations
\begin{itemize}
	\item $  d_*U $ is the normalized Haar measure on $\mathbb{\mathbb{U}}(N)$: $\int_{\mathbb{U}(N)}d_*U =1$
	
	\item $Z$ is a complex matrix
	$$
	d\Omega(Z,Z^\dag)  =\,\pi^{-n^2}\,\text{e}^{-\tr \left(ZZ^\dag\right)}\,
	\prod_{i,j=1}^N \,d \Re Z_{ij}d \Im Z_{ij}
	$$

	\item Let $M$ be a Hermitian matrix the measure is defined
	$$
	dM= \, \prod_{i\le j}
	d\Re M_{ij} \prod_{i<j} d\Im M
	$$	
\end{itemize}
It is known \cite{Mac}
\be\label{s-s-N_lambda-1}
\int s_\lambda(Z)s_\mu(Z^\dag)\,d\Omega(Z,Z^\dag) = (N)_\lambda\delta_{\lambda,\mu}
\ee
where $(N)_\lambda:=\prod_{(i.j)\in\lambda}(N+j-i)$ is the Pochhammer symbol
related to $\lambda$. A similar relation
was used in \cite{O-Acta},\cite{HO-2006},\cite{O-2004-New},\cite{AMMN-2014},\cite{OShiota-2004}, for models of Hermitian, complex
and normal matrices.

By $\mathbb{I}_N$ we shall denote the $N\times N$ identity matrix.
We  recall that
$$ s_\lambda(\mathbb{I}_N)=(N)_\lambda s_\lambda(\bpow_\infty)\,,
\qquad s_\lambda(\bpow_\infty) = \frac{\operatorname{dim}\lambda}{d!},\quad d=|\lambda|$$.
\end{exampB}
\begin{exampB}
{\bf Three branch points.}
The generating function for $\mathbb{RP}^2$ Hurwitz numbers with three ramification points, having three
arbitrary profiles:
\be\label{3-points-integral}
\sum_{\lambda,\,\ell(\lambda) \le N}
\frac{s_\lambda(\bpow^{(1)}) s_\lambda(\Lambda) s_\lambda(\bpow^{(2)})}{\left( s_\lambda(\bpow_\infty) \right)^2}
\ee
\[
= \,\int \,\tau^{\rm B}_1\left( Z_1 \Lambda Z_2 \right)  \,\prod_{i=1,2} \,
\text{e}^{\tr \texttt{V}( Z^\dag_i,\,\bpow^{(i)})}\,d\Omega(Z_i,Z^\dag_i)
\]
If $\bpow^{(2)}=\bpow(\texttt{q},\texttt{t})$ with any given parameters $\texttt{q},\texttt{t}$, and $\Lambda=\mathbb{I}_N$
then (\ref{3-points-integral}) is the hypergeometric BKP tau function. 
\end{exampB}
\begin{exampB}
{\bf `Projective' Hermitian two-matrix model}.
The following integral
\[
\int \tau^{\rm B}_1(c M_2)  \text{e}^{\tr \texttt{V}(M_1,\bpow)+\tr (M_1 M_2)}dM_1dM_2 =
\sum_{\lambda}\,c^{|\lambda|} (N)_\lambda  s_\lambda(\bpow)
\]
where $M_1,M_2$ are Hermitian matrices is an example of the hypergeometric BKP tau function.
\end{exampB}
\begin{exampB}
{\bf Unitary matrices.} Generating series for projective Hurwitz numbers with arbitrary profiles
in $n$ branch points and restricted profiles in other points:
\begin{multline}
\label{multimatrix-unitary-RP2'}
\int \text{e}^{\tr (c U_1^\dag \dots U_{n+m}^\dag)}
\left(\prod_{i=n+1}^{n+m} \tau^{\rm B}_1(U_i) d_*U_i \right)
\left(\prod_{i=1}^{n} \tau^{\rm KP}_1(U_i,\bpow^{(i)}) d_*U_i \right)\\
=
\sum_{d\ge 0}c^d \left( d! \right)^{1-m} \sum_{\lambda,\, |\lambda|=
	d\atop \ell(\lambda)\le N}\, \left(\frac{\operatorname{dim}\lambda}{d!}  \right)^{2-m}
\left(\frac{s_\lambda(\mathbb{I}_N)}{\operatorname{dim} \lambda} \right)^{1-m-n}
\prod_{i=1}^n \frac{s_\lambda(\bpow^{(i)})}{\operatorname{dim} \lambda}
\end{multline}

Here $\bpow^{(i)}$ are parameters. This series generate certain linear combination of Hurwitz numbers for base surfaces
with Euler characteristic $2-m,\,m\ge 0$. In case $n=1$ this BKP tau function may be viewed as an analogue of the generating function of
the so-called non-connected Bousquet-Melou-Schaeffer numbers
(see Example 2.16 in \cite{KazarianLando}).
In case $n=m=1$ we obtain the following BKP tau function
\[
\int \tau^{\rm B}_1(U_2)  \text{e}^{\tr \texttt{V}(U_1,\bpow)+\tr (cU_1^\dag U_2^\dag)}d_*U_1d_*U_2 =
\sum_{\lambda\atop \ell(\lambda)\le N}\,c^{|\lambda|}\frac{s_\lambda(\bpow)}{(N)_\lambda}
\]
\end{exampB}
\begin{exampB}
{\bf Integrals over complex matrices}.
A pair of examples. 
An analogue of Belyi curves generating function \cite{Zog},\cite{Chekhov-2014} is as follows:
\begin{multline}
\sum_{l=1}^N N^l\sum_{ \Delta^{(1)},\dots,\Delta^{(n+1)}\atop \ell(\Delta^{n+1})=l} c^d
H_{\textsc{e}}(d;\Delta^{(1)},\dots,\Delta^{(n+1)})
\prod_{i=1}^{n} \bpow^{(i)}_{\Delta^{(i)}}\\
=\sum_{\lambda}c^{|\lambda|}\frac{(d!)^{m-2}(N)_\lambda}{(\operatorname{dim}\lambda)^{m-2}}\,
\prod_{i=1}^{n}\frac{s_\lambda(\bpow^{(i)})}{s_\lambda(\bpow_\infty)}\\
=\int \text{e}^{\tr (cZ_1^\dag \dots Z_{n+m}^\dag)}
\left(\prod_{i=n+1}^{n+m} \tau^{\rm B}_1(Z_i) d\Omega(Z_i,Z_i^\dag) \right)\\
\times\left(\prod_{i=1}^{n} \tau^{\rm KP}_1(Z_i,\bpow^{(i)}) d\Omega(Z_i,Z_i^\dag) \right)
\end{multline}
where $\textsc{e}=2-m$ is the Euler characteristic of the base surface.

The series in the following example generates the projective Hurwitz numbers themselves where to get rid
of the factor $(N)_\lambda$ in the sum over partitions we use mixed integration over $\mathbb{U}(N)$ and over
complex matrices:
\begin{multline}
\sum_{ \Delta^{(1)},\dots,\Delta^{(n)}}\, c^d\,
H_{1}(d;\Delta^{(1)},\dots,\Delta^{(n)})\,
\prod_{i=1}^{n} \bpow^{(i)}_{\Delta^{(i)}}\\
=\sum_{\lambda,\,\ell(\lambda)\le N}\,c^{|\lambda|} \frac{\operatorname{dim}\lambda}{d!}\,
\prod_{i=1}^{n}\frac{s_\lambda(\bpow^{(i)})}{s_\lambda(\bpow_\infty)}\\
=\,\int \tau_1^{\rm KP}(c U^\dag Z_1^\dag \cdots Z_k^\dag,\bpow^{(n)})\tau_1^{\rm B}(U)d_*U \prod_{i=1}^{n-1}
\tau_1^{\rm KP}(Z_i,\bpow^{(i)}) d\Omega(Z_i,Z_i^\dag)
\end{multline}
Here $Z,Z_i,\,i=1,\dots,n-1$ are complex $N\times N$ matrices and $U\in\mathbb{U}(N)$. As in the previous examples
one can specify all sets $\bpow^{(i)}=\bpow(\texttt{q}_i,\texttt{t}_i),\,i=1,\dots,n$ except a single one which in 
this case has the meaning of the BKP higher times.
\end{exampB}

\section{The unitary ensemble as an example of a tensor model and Hurwitz numbers}

\subsection{One matrix model and combinatorics of graphs}
Let me recall some facts about Dyson-Wigner unitary ensemble and one-matrix model.
The probability measure on the space of $N\times N$ Hermitian matrices is defined as
\be\label{DWEns-measure}
d\nu_N(h)=c_N\prod_{i>j} e^{-\left(\Re h_{ij}\right)^2-\left( \Im h_{ij} \right)^2  }
d\Re h_{ij} d\Im h_{ij}\prod_{i=1}^N d h_{ii}
\ee
see \cite{Mehta}, the constant $c_N$ is chosen from the condition $\int d\nu_N(h)=1$ where one integrates
over the space of $N\times N$ Hemitian matrices. The expectation value for the Dyson-Wigner ensemble is defined as
\[
\mathbb{E}^{DW}_N(f)=\int f(h) d\nu_N(h)
\]
The famous pioner works of Kazakov, Brezin \cite{BrezinKazakov}, Migdal and Gross \cite{MigdalGross} 
relates this model to 
the theory of the two-dimensional quantum gravity and combinatorial models of
Riemann surfaces on the one hand and to the Painleve equation to the other hand.
The relation to the Virasoro constrainted tau functions of the Toda lattice was worked out
in \cite{GMMMO}.

Here we review the combinatorial aspects of this model in very short. For details I send the reader
to the bright review of this topic in \cite{ZL}.
Consider the following expectaion value
\be\label{DW-spectral-inv}
\mathbb{E}^{DW}_N\left(\tr h^{\lambda_1}\cdots \tr h^{\lambda_\ell}  \right)=:
\mathbb{E}^{DW}_N\left({\bf p}_\lambda(h)\right)
\ee
where $\lambda=(\lambda_1,\dots,\lambda_\ell)$ is a partition of length $\ell$ (it means that 
$\lambda_\ell >0$). One can check that this expectation value vanishes if the weight 
$|\lambda|=\lambda_1+\cdots +\lambda_\ell$ if the
partition $\lambda$ is odd. Let $|\lambda|=2n$.
has the following meaning. Let us consider  $\ell$ polygons
with resectively $\lambda_1,\, \lambda_2,\,\dots ,\,\lambda_\ell$ edges. We imply that the polygons
are (say, clockwise) oriented. Each edge is linked with a single edge. Let us connect such pairs
by a line - as we did it before in subsection \ref{Product of complex...}. We will call these lines which connect edges
of the same polygon chords, and  lines which connect different polygon links.
One can glue all edges connected (either by chord or by link) in the pairwise way, identifying
the end of one edge with the beginning of the other one (we remember that polygons are oriented). 

The central statement is
that the expectation (\ref{DW-spectral-inv}) counts the number of the ways one can glue the polygons, 
see for instance Chapter 3.3.called "Matrix Integrals for Multiface maps"  in \cite{ZL} for the best
review. Each way 
of gluing yields the model of orientable two-dimentional
surface $\Sigma_{g^*}$ of genus $g^*$ and the ribbon graph with $n$ edges and with 
$v=n-\ell+g^*$ vertices. 

The expectations (\ref{DW-spectral-inv}) are generated by the famous one-matrix model, introduced
in \cite{BrezinKazakov}:
\be\label{1MM}
\mathbb{E}^{DW}_N\left(e^{N\tr \texttt{V}(h,{\bf p})} \right)=\sum_{\lambda} \frac{1}{z_\lambda}N^{\ell(\lambda)}
\mathbb{E}^{DW}_N\left({\bf p}_\lambda(h)\right){\bf p}_\lambda
\ee
where ${\bf p}=(p_1,p_2,\dots)$ are parameters (the coupling constants).\footnote{In the original model all $p_i=0$ except
$p_2$ and $p_3$, the infinite set of parameters - Toda lattice higher times - was introduced in
\cite{GMMMO}.} 

To get the statement one need to do the following steps

(1) to write down each trace, say, $\tr h^{k}$ as $S_k=h_{i_1,i_2}h_{i_2,i_3}\cdots h_{i_k,i_1}$ where
we imply the summation over repeated indices.  We assign a $k$-
polygon to each trace $\tr h^k$, thus, we get $\ell$ polygons respectively of sizes 
$\lambda_1,\dots,\lambda_\ell$. Each term in the sum $S_k$ is labeled by a given set
$i_1,\dots,i_k$ which labels vertices of the polygon in, say, anti-clockwise direction,
while the edge between the vertices $i_a,i_{a+1}$ are assigned to the entry $h_{i_a,i_{a+1}}$.

(2) Consider $\mathbb{E}^{DW}_N(S_{\lambda_1}\cdots S_{\lambda_\ell})$ and
 take Gauss integrals of each term in the sum over all variables. Then, only these terms contribute
whose all $k$ factors meet their pair.
One uses the chord diagrams to denote the Wick's pairing of the entries.
Each chord connects a pair of either of $h_{ij}$ and $h_{ji}$ where $N\ge i>j$, or the pairing of $h_{ii}$
with itself where $i=1,\dots,N$. The pairing means gluing of the sides of polygons. One gets
the oriented two-dimensional surface of a genus which is the genus of the chord diagram $g^*$

(3) the result of the Gauss integration of
each monomial term of the product $S_{\lambda_1}\cdots S_{\lambda_\ell}$ is equal either 1 or 0.
Thus, the whole sum (\ref{1MM}) is equal to the number of possible chord diagrams up to
the weight of the automorphism group of each chord diagram (not to count it twice or more times).

\subsection{A tensor model based on the one-matrix model}

Consider the $N\times N$ matrix $h$ with noncommuting entries. In our case,
one can think of the Hermitian matrix $H$ (of the size $L\times L$ where $L=N M$)  splitted
into blocks of the size $N\times N$. Then, each entry may be labeled by 4 indices $h_{i,j}^{a,b}$
where $i,j=1,\dots,N$ and $a,b=1,\dots, M$. Let us consider Hermitian $H$. Then $h_{ij}=h_{ji}^\dag$.
Let us introduce axillary 
complex matrices $Z_i,\,i=1,\dots,N$ such that 
$h_{ii}=\frac{1}{\sqrt{2}}\left(Z_i+Z_i^\dag \right)$
(of cause, such matrices are not defined in the unique way) and introduce 
$y_{i}:=\frac{1}{\sqrt{-2}}(Z_i-Z_i^\dag)$ which is Hermitian.

Consider Dydson-Wigner unitary ensemble of the $L\times L$ matrices $H$. The probability measure
can be written as
\be
d\nu_L(H)= \prod_{i>j}^N d\mu\left(h_{ij}\right)\prod_{i=1}^N d\nu(h_{ii})\int \prod_{i=1}^N d\nu_N(y_i)
=\prod_{i>j}^N d\mu\left(h_{ij}\right)\prod_{i=1}^N d\mu(Z_i)
\ee
where the measures $d\mu$ and $d\nu$ are defined respectively by (\ref{CGEns-measure}) and by 
(\ref{DWEns-measure}).

On the other hand, it is the model of $\frac 12 N(N+1)$ independent complex Ginibre ensembles
without sources (all sources are identity matrices). 
This is the ensembles of complex matrices $\{h_{i,j},\,N\ge i>j \}$ and $\{Z_i,\,i=1,\dots,N\}$.
We should keep in mind that the set of matrices
$Z_i,\,i=1,\dots,N$ enters into
\be
\mathbb{E}_L\left( \tr_3 H^{\lambda_1}\cdots \tr_3 H^{\lambda_\ell} \right)
\ee
slightly differenetly.

Let us consider the one-matrix model based on $L\times L$ matrices which is known to be
the Virasoro constraint 1D Toda lattice (which is
also a special KP, 2-KP and also 2D Toda lattice tau function):
\be
\tau_L({\bf p})=\mathbb{E}^{DW}_L\left( e^{\tr \texttt{V}(H,{\bf p})} \right)=
\sum_\lambda \left( L\right)_\lambda s_\lambda({\bf p})s_\lambda(0,1,0,\dots)
\ee

Let us consider the same products of $S_k=h_{i_1,i_2}h_{i_2,i_3}\cdots h_{i_k,i_1}$ but
now

At last, let us note that the measure $d\nu_L(H)$ can be treated
as the measure of the simple tensor model written as
\be
d\omega(h)= e^{-\sum_{N\ge i\ge j,\atop a, b=1,\dots,M} h^{a,b}_{i,j}h^{b,a}_{j,i}}\times
\ee
\[
\prod_{N\ge i > j,\atop a, b=1,\dots,M} 
d\Re h_{i,j}^{a,b} d\Im h_{i,j}^{a,b}\prod_{i=1,\dots,N\atop M\ge a > b}
d\Re h_{ii}^{a,b}d\Im h_{ii}^{a,b}\prod_{i,a} d h_{i,i}^{a,a}
\]

\end{document}